\documentclass{article}

\usepackage{arxiv}

\usepackage[utf8]{inputenc} % allow utf-8 input
\usepackage[T1]{fontenc}    % use 8-bit T1 fonts
\usepackage{hyperref}       % hyperlinks
\usepackage{url}            % simple URL typesetting
\usepackage{booktabs}       % professional-quality tables
\usepackage{amsfonts}       % blackboard math symbols
\usepackage{nicefrac}       % compact symbols for 1/2, etc.
\usepackage{microtype}      % microtypography
\usepackage{lipsum}		% Can be removed after putting your text content
\usepackage{graphicx, subfigure}
\usepackage{natbib}
\usepackage{doi}

\title{BEAMERS: Brain-Engaged, Active Music-based Emotion Regulation System}

\author{{\includegraphics[scale=0.06]{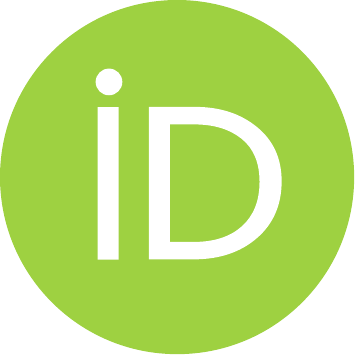}\hspace{1mm}Jiyang Li} \\
	Department of Computer Science and Engineering\\
	University at Buffalo, State University of New York\\
	Buffalo, NY, 14260, USA \\
	\texttt{jiyangli@buffalo.edu} \\
	\And
	{\includegraphics[scale=0.06]{orcid.pdf}\hspace{1mm}Wei Wang} \\
	Department of Computer Science and Engineering\\\
	University at Buffalo, State University of New York\\
	Buffalo, NY, 14260, USA \\
	\texttt{wwang49@buffalo.edu} \\
    \And
	{\includegraphics[scale=0.06]{orcid.pdf}\hspace{1mm}Kratika Bhagtani} \\
	Department of Electrical Engineering\\\
	Indian Institute of Technology Gandhinagar\\
	Gandhinagar, Gujarat, India \\
	\texttt{kratika.bhagtani@alumni.iitgn.ac.in} \\
    \And
	{\includegraphics[scale=0.06]{orcid.pdf}\hspace{1mm}Yincheng Jin} \\
	Department of Computer Science and Engineering\\\
	University at Buffalo, State University of New York\\
	Buffalo, NY, 14260, USA \\
	\texttt{yincheng@buffalo.edu} \\    
    \And
	{\includegraphics[scale=0.06]{orcid.pdf}\hspace{1mm}Zhanpeng Jin} \\
	Department of Computer Science and Engineering\\\
	University at Buffalo, State University of New York\\
	Buffalo, NY, 14260, USA \\
	\texttt{zjin@buffalo.edu} \\
}

\date{}

\renewcommand{\shorttitle}{BEAMERS: Brain-Engaged, Active Music-based Emotion Regulation System}

\hypersetup{
pdftitle={A template for the arxiv style},
pdfsubject={q-bio.NC, q-bio.QM},
pdfauthor={David S.~Hippocampus, Elias D.~Striatum},
pdfkeywords={First keyword, Second keyword, More},
}

\begin{document}
\maketitle

\begin{abstract}
With the increasing demands of emotion comprehension and regulation in our daily life, a customized music-based emotion regulation system is introduced by employing current EEG information and song features, which predicts users' emotion variation in the valence-arousal model before recommending music. The work shows that: (1) a novel music-based emotion regulation system with a commercial EEG device is designed without employing deterministic emotion recognition models for daily usage; (2) the system considers users' variant emotions towards the same song, and by which calculate user's emotion instability and it is in accordance with Big Five Personality Test; (3) the system supports different emotion regulation styles with users' designation of desired emotion variation, and achieves an accuracy of over $0.85$ with 2-seconds EEG data; (4) people feel easier to report their emotion variation comparing with absolute emotional states, and would accept a more delicate music recommendation system for emotion regulation according to the questionnaire.
\end{abstract}

% keywords can be removed
\keywords{Human emotion \and emotion regulation \and music recommendation \and emotion instability}

\section{Introduction}
Given dramatic changes in lifestyles in modern society, millions of people nowadays are affected by anxiety, depression, exhaustion, and other emotional problems. The demand for proper emotional care is accumulating and accelerating at an increasingly rapid rate. Music, as being extensively studied in literature \cite{Koelsch2014, Taruffi2017, Siddharth2019, Ehrlich2019}, is proven to be one of the most effective and accessible manners to evoke emotions and influence moods \cite{goldstein1980thrills}. And it has been seen and reported that people are spending more time listening to music over the past decades with the development of portable music players and streaming platforms like YouTube, Spotify, and iTunes \cite{AudienceNet2018, Music3602017}. Inevitably, music shapes our life in various ways, including helping people better focus their attention, releasing more dopamine for a better mood, exercising with more energy, and boosting our creativity. 

People's sensation towards a music piece is subtle and might be distinct every time he/she listens to it. Biometric signals are collected along with people's self-reporting to understand human emotional reactions towards music. With the development of less obtrusive, non-invasive EEG devices, EEG becomes the dominant modality for studying brain activities, including emotion recognition in human-computer interactions (HCI) studies \cite{mihajlovic2014wearable}. A large number of studies have explored the music-induced neural correlates of perception and cognition, and provided theoretical supports for the application of music-based emotion regulation. People are looking forward to the day when wearable EEG devices could bridge the gap between the theoretical studies and the applications that can benefit users in their daily life. However, more accessible, easy-to-use, and interactive approaches for effective emotion regulation and mental health are still largely under-explored.

% it can be used in our daily life for example, the emotion regulation purpose or personal recommendation systems.  people are looking forward that Noninvasive surface EEG is. would become possible for real-life usage like emotion regulation. However, there is no such application actually being used in our daily life.

% based on Magnetic Resonance Imaging (fMRI), Positron Emission Tomography (PET), magnetoencephalography (MEG) and EEG \cite{moore2013systematic}

% \begin{figure}[tbp!]
%   \centering
%   \includegraphics[width=0.7\columnwidth]{figures/scenario.pdf}
%   \caption{Two scenarios that when people try to regulate their emotions by listening to music. Scenario 1 is the commercial music platform we use daily. Scenario 2 is the previous work that uses EEG as an indicator of emotional state to choose music pieces for emotion regulation.}
%   \label{scenario}
% \end{figure}

\begin{figure*}[tbp!]
  \centering
  \includegraphics[width=\columnwidth]{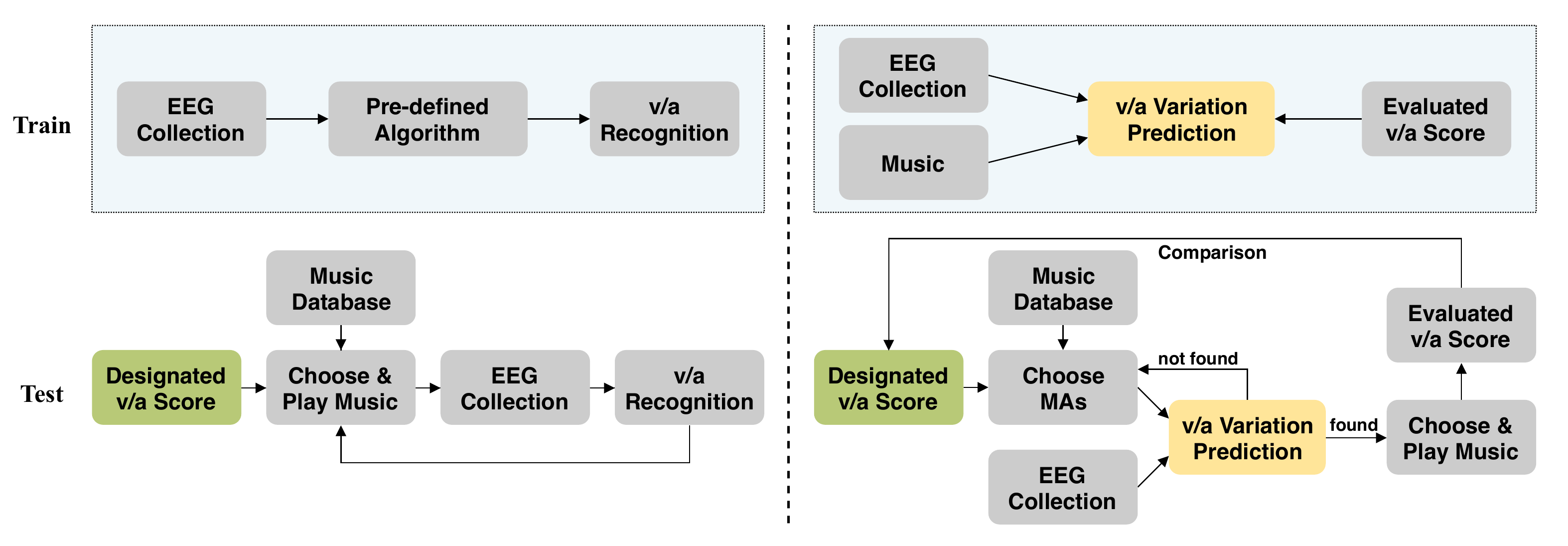}
  \caption{Workflow comparison of existing systems based on the emotion recognition model (left) and our proposed system (right). The upper framework in the block is the training process, and the lower framework is the testing process. MAs is the acronym for Music Alternatives.}
  \label{compare}
\end{figure*}

% They differ from the input channels of EEG signal, the emotion model and classification model, music selection strategies etc. 

The introduction of EEG-based emotion recognition models makes it possible that computers can be empathetic. Existing studies are primarily based on emotion recognition models, as shown on the left of Figure \ref{compare}. Some of them employ definitive approaches without users' self-reporting to provide user-determined feedback \cite{ali2016eeg, Adamos2016, Ramirez2018, sourina2012real, moore2013systematic}. However, the emotion recognition algorithms vary widely and will reach different interpretations for similar data \cite{reed2017learning, yoon2017comparison}. Users would over-accept and defer to the feedback, even if it contradicts with users’ own interpretations \cite{hollis2018being}. Another drawback is the limited music choices. To work with the emotion recognition model, the system needs to label the music pieces based on users' evoked emotions. With the limited emotional states, a large music library would make their selection strategy ineffective and inefficient. And it's unfeasible to ask users to listen to a large number of songs to acquire labels. Some research asked users or professional music therapists to select music before the experiment. Other studies labeled music by the emotion expressed in music (``music emotion'') \cite{jun2010music, lu2009novel, Juslin2004}, and then recommended a specific song by matching its emotion to the detected user's emotional state. Such a strategy assumes that the music-listening preferences and styles of an individual are always consistent with his/her emotional state at the moment. However, the music-listening styles of users are so varying \cite{ferwerda2014enhancing}, and their emotional reactions change over time and under different situations \cite{thoma2012emotion}. And more importantly, emotions that a listener experiences are distinct from the judgments of what emotion is expressed in music \cite{lewis2010handbook}, which was proved by the low match rate in our study. 

% by the emotion recognition model, and make the decision of recommended music pieces by matching the detected emotional state.

Most existing research uses discrete emotion models with states like `happy', `angry', and `sad', or continuous models like Russell's circumplex model \cite{russell1980circumplex} for emotion evaluation. One problem is that systems relying on the emotion recognition models assume that the stimulus would elicit emotions effectively \cite{alarcao2017emotions}. Thus when the emotion is not successfully elicited, self-assessment would be tough, and the performance of such systems would be questionable. A more essential problem is that the detected emotional states only act as a reference and don't link the music and users in a relation. Under such circumstances, users can regulate emotions themselves by searching a playlist with a certain `mood' on streaming platforms and switching songs referring to their emotional states. Given all the limitations and shortcomings above, it is argued that the conventional emotion recognition models are insufficient and unreliable for the design of emotion regulation applications. 

% Under such circumstances, advanced technologies are not required, instead, users can listen to music as usual under certain guidance or ask for music therapists' help.
% Besides, the decision of the next song is independent from the interaction between user and music but determined by matching the recognized user's emotion and the emotion expressed in music as shown on the left of \ref{compare}. is necessary for studying emotion-related brain mechanisms but

% Music streaming platforms can learn users' preferences of music, but it is insufficient for emotion regulation as the preferred song could induce unwanted emotion. EEG-based systems can recognize users' emotional state but it doesn't mean that the music could be selected properly. 

To design an individual-oriented, truly personalized music-based emotion regulation application, we need to ask and address two questions: (1) Will the user like the chosen/recommended music? (2) Will this music emotionally affect the user in a way he/she wants? Although some existing music recommendation approaches have partially addressed these two questions by collecting information about users' profiles and preferences \cite{Hu2011, Schedl2018}, there is still one essential question to be considered in terms of the complex and diverse variations of the user's emotion over time: What is the user's current emotional state and what will his/her emotional state be after listening to this song? With the advances of ubiquitous HCI technologies, it is necessary and also becomes possible to address these questions for understanding and handling human's ever-changing emotions.

% Some neuro-feedback music therapy studies require users to pre-select the preferred musics \cite{ramirez2015musical, ramirez2018eeg} to avoid unsatisfied or unfavorable music choices, but it is inconvenient when using in the daily life.

% Most of papers differ in the emotion recognition method and the strategy of choosing music. [] Another problem is that those systems recognize the user's emotions after listening to the recommended song, of the previously recommended music piece and highly relies on the empirical study of emotion recognition algorithms for  The independence between users' EEG data and the music selection strategies recommendation has significant influence on the capability of learning the user's music preference.

% Here we propose that the decision should base on the user's present emotional state besides side information to match `the moment' of a user. we try to figure out a method that doesn't employ   of doesn't depend on the emotion recognition model and overcomes its drawbacks, as well as answers questions addressed above.

To this aim, in this study, we propose an emotion regulation system that predicts the emotion variation (i.e., the emotion change after listening to a song compared with before) before any recommendation. In this system, the song and the user's current emotional state are two factors that decide how his/her emotion is going to be changed after listening to this song. Therefore, two independent variables: the music and the current EEG information, define and indicate the consequential emotion variation, as shown on the right of Figure \ref{compare}. Concatenations of each song selected in music alternatives (MAs) and the current EEG information are fed into the prediction model sequentially, then a music piece is chosen from MAs based on their predicted emotion variations. 

%Emotiv EPOC+ wearable headset is chosen from Neurosky, CGX, Muse after we reviewing the off-the-shelf commercial EEG devices to collect users' EEG data, which is one of the most wearable EEG devices and proven to be effective for mood induction studies \cite{rodriguez2013validation}. 

% that is represented by valence and arousal coordinates as shown in Figure~\ref{Russell} in our study. 

% Here the input of music and EEG information are their extracted features. 

% Unlike the absolute emotional state, emotion variation describes the change of emotion compared with the last state, 

% And Emotiv EPOC+ with 14 channels is chosen to collect users' EEG data, which is one of the most wearable EEG devices in the market and proven to be effective for mood induction studies \cite{rodriguez2013validation}.   

\begin{figure}[tbp!]
  \centering
  \includegraphics[width=0.5\columnwidth]{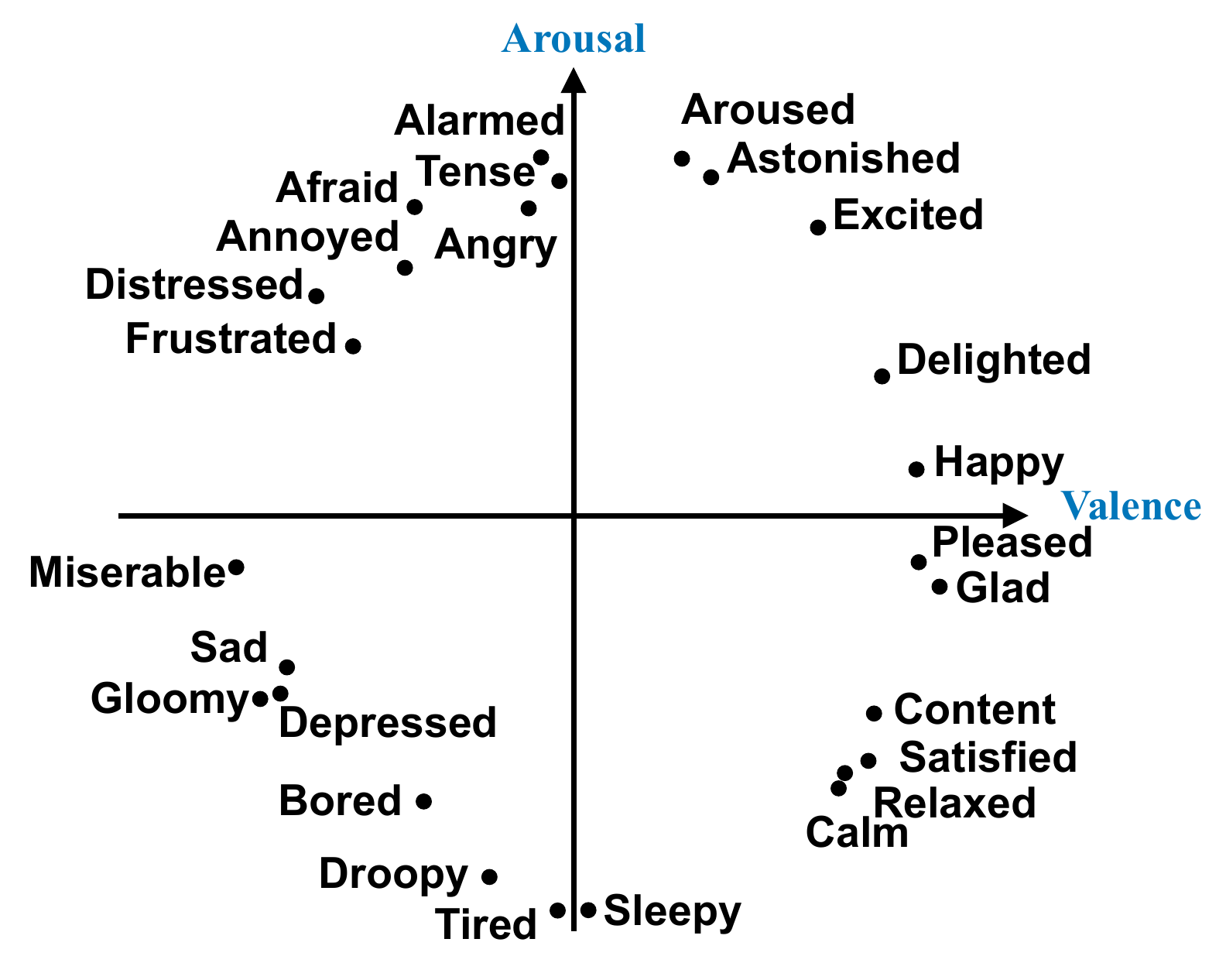}
  \caption{Russell's circumplex model of affect \protect\cite{russell1980circumplex}. Locations of affects on the model correspond to their valence and arousal values. The y-axis represents arousal (intensity) and the x-axis represents valence.}
  \label{Russell}
  \vspace{-15pt}
\end{figure}

Unlike the definitive emotional state, emotion variation describes how much the emotion is changed. That means, it will be determined based on an individual's current emotional state, and will be affected or induced by emotional stimuli (listening to a song in our case). It requires less effort for users to report even when the emotion sustains, which is ``neutral'' or a close-to-zero value. And the model is able to deal with the situation that ``I'm feeling better but still sad,'' which is a positive variation but not a ``joyful'' state after listening to a song. Discrete emotion models with the classes like ``happy'', ``angry'', and ``sad’' are not suitable for the assessment of emotional variations. Thus we apply Russell's circumplex model with valence and arousal coordinates (Figure~\ref{Russell}) for users' self-assessment, which is called valence-arousal (v/a) scores in this study. Yet, the challenge is that, continuous values of emotion variation are not realistic or feasible for models to learn; instead, discrete emotion classes are learned to represent people's typical feelings \cite{lin2010eeg, baumgartner2006emotion}. Thus, for the proposed emotion variation prediction model, we seek to simplify users' continuous v/a scores into four classes as the four quadrants in the valence-arousal model by positive/negative values, by which we predict the direction of emotion variations instead of the absolute variation amplitude. The training process concatenates EEG features with song features to train the binary valence and arousal models offline. In the testing process, a user needs to first designate his/her desired emotion variation, and their current EEG information will be collected. Meanwhile, music alternatives (MAs) are coarsely selected to narrow down songs for the prediction model, and are combined with EEG information to predict the direction of the user's emotional variation on the fly. Finally, a song from MAs will be chosen if the predicted class of v/a scores matches the user's designated v/a score.

% our system is incapable And it has received great support in studies of emotions, cross-culture comparison and psychometric studies \cite{posner2005circumplex}. The same model is also employed to conceptualize emotion expressed in music in Music Information Retrieval (MIR) studies \cite{soleymani20131000}. 

%  Dimensional models are represented by continuous coordinates, the most famous dimensional model is the circumplex model \cite{russell1980circumplex} represented by valence and arousal. Circumplex model has received great support in studies of emotions, cross-culture comparison and psychometric studies \cite{posner2005circumplex}. Affects represented by four combinations of these two bipolar coordinates can be induced by musical pieces \cite{rickard2004intense, vieillard2008happy}, and show unique neural patterns \cite{altenmuller2002hits}.
 
% Thus we assume that the user's emotion variation is always decided by his/her presented emotional state and the perceived song.  also insufficient for the lack of proper music recommendation strategies.

% (which narrows down the input music pieces to alleviate the computational workload of the prediction model)

% In addition, the determined emotion recognition algorithm is not employed. Instead, we will develop a model to predict the user's emotion variations towards the song based on previous interactive information, then recommend the proper song to achieve the purpose of emotion regulation. 

% In this paper, we propose an EEG-guided music recommendation system that looks ahead to prognosticate the user's emotion variations towards music to make the proper recommendation. We assume that the primary affect-inducing interventions that decide the user's emotion variations are the user's present emotional state and the perceived music. 

% to concatenate with the present EEG respectively . Then the customized model predicts the user's v/a variation for each pair of EEG-song and compared it with the designated emotional state, and chooses the matched song for recommendation. 

%  are concatenated one-to-one (EEG-song)
 
In this paper, we aim to break the limitations of existing music-based emotion regulation systems and make the following contributions: (1) our approach bridges the gap between the theoretical studies and the practical and usable interactive systems for daily usage by proposing a dynamic emotion variation model, instead of using the conventional definitive emotion recognition methods; (2) based on the qualitative prediction of emotion variations, our system is able to recommend proper songs, being optimized towards both users' listening preferences and their desired change of emotion; and (3) the user's different emotion variations resulting from one song can be distinguished by the system, by which we evaluate the user's emotion instability and imply that it could be utilized as an indicator for mental health-related applications. Lastly, a simple questionnaire was developed and conducted to evaluate the user acceptance and usability of the proposed system.

\section{Related Work}
\subsection{Emotion Models}
In affective science studies, the terms of affect, emotion, and mood are precisely differentiated. Affect is not directed at a specific event and usually lasts for a short time. Emotion is intense and directed, which is stimulated by some events or objects and also lasts for a short time. Mood lasts longer and is less intense \cite{liu2017many}. Unlike affective science studies, our study leaves aside the debate of whether emotions should be represented as a point in a dimensional valence-arousal space, as well as the distinction between affect, emotion, and mood.  

% which focuses more on the phenomena such as feelings, moods, attitudes, affective styles \cite{calvo2010affect}

Researchers introduced various emotion models to conceptualize human emotion, among which discrete and dimensional models are employed mostly \cite{eerola2013review}. Discrete models, relating to the theory of basic emotions \cite{ekman1992argument}, have been widely adopted because their concepts are easily explained and discriminated. However, studies, even including all variants of basic emotions, can hardly cover the whole range of human emotions. And some of the categories can be musically inappropriate when evaluating the emotion evoked by music or emotion expressed by music \cite{balkwill1999cross}. Dimensional models are represented by continuous coordinates, the most famous of which is the circumplex model \cite{russell1980circumplex} represented by valence and arousal. The circumplex model has received significant support in studies of emotions, cross-culture comparison, and psychometric studies \cite{posner2005circumplex}. Affects represented by four combinations of these two bipolar coordinates can be induced by musical pieces \cite{rickard2004intense, vieillard2008happy}, and show unique neural patterns \cite{altenmuller2002hits}. Because of a more inclusive representation of emotions, the dimensional model has been extensively used in the Affective Computing (AC) research community

% In the interdisciplinary field of Affective Computing (AC), researchers introduced emotion models with discrete types such as surprise, joy, rage, fear, disgust and so on \cite{nowlis1956description, tomkins1962affect}, as well as continuous models with two-dimensional valence-arousal (v-a) model \cite{russell1980circumplex}, or three-dimensional pleasure-arousal-dominance (PAD) model \cite{mehrabian1980basic}. 

\subsection{Music Emotion}
The inherent emotional state of an acoustic event (``music emotion'' \cite{Schyff2017}) is conveyed by a composer while writing it or a musician while performing it \cite{weninger2013acoustics}. And it can be classified by low-level descriptors (LLDs), such as zero-crossing rate, Mel-frequency cepstral coefficients (MFCC), spectral roll-off, and so on. Extensive prior studies have been performed and dedicated to the development of automatic Music Information Retrieval (MIR) systems. Functions for calculating acoustic features were packed in toolboxes such as the MIR toolbox \cite{lartillot2007matlab}, openSMILE \cite{eyben2013recent}, and librosa \cite{mcfee2015librosa}. 

Databases with music emotion have been developed for MIR competitions, such as the annual music information retrieval evaluation exchange (MIREX) that started in 2007 \cite{downie20082007}. However, they introduced five discrete emotions in the task instead of assigning a continuous affect model. For a better understanding of the music effects psychologically, we referred to the 1000 songs database \cite{soleymani20131000}, where the annotations of all music emotions were done by crowdsourcing on the Amazon Mechanical Turk, based on Russell's valence-arousal model. Low-quality and unqualified annotators have been removed to guarantee the quality of all annotations.

% which was a great advantage for our research using the same model to represent listeners' emotions.

\subsection{EEG-based Emotion Recognition}
% Emotion of human (emphasize the EEG)
Recognizing people's extensively varying induced emotions is demanding and challenging. A variety of physical and physiological information have been employed to estimate human emotions in the past two decades, such as facial expressions \cite{black1995tracking,essa1997coding}, heartbeats \cite{valenza2013nonlinear}, speech \cite{dellaert1996recognizing, nwe2001speech}, body gestures \cite{shibata2012emotion}, and EEG \cite{takahashi2004remarks, lin2010eeg, alzoubi2009classification}. EEG is a non-invasive electrophysiological technique that uses small electrodes placed on the scalp to record brainwave patterns. With the advances in Brain-Computer Interface (BCI) technologies, EEG signals can be collected from wearable devices with one to hundreds of channels \cite{gui2019survey}. For example, Emotiv EPOC+ neuro-headset with 14 channels was proven to be user-friendly and effective for mood induction studies \cite{rodriguez2013validation}. 

Nowadays, it has been well studied and recommended that, as a common practice, the power spectra of the EEG can be divided into five bands --- delta ($\delta$: 1-3 Hz), theta ($\theta$: 4-7 Hz), alpha ($\alpha$: 8-13 Hz), beta ($\beta$: 14-30 Hz), and gamma ($\gamma$: 31-50 Hz) \cite{mantini2007electrophysiological}. The band-related brain activities have been explored to represent various emotions \cite{dasdemir2017analysis}, including the asymmetry at the anterior areas of the brain \cite{schmidt2001frontal}. Self-reporting is still the most traditional and common way to obtain the subject's emotional states; but it has been long questioned because of the subjectivity and instability of each individual viewpoint. Self-Assessment Manikin (SAM) \cite{bradley1994measuring} was highlighted to assist the self-assessment of emotions in some projects \cite{nie2011eeg, daly2015identifying, khosrowabadi2010eeg}. However, it is still challenging for most subjects to accurately assess and label their definitive emotional states. Instead, emotion variations -- how does an event or stimulus change the emotion from the last state, either positively or negatively --- is easier for subjects to evaluate and describe, which hence are selected as the benchmark in our study.  

% Previous paper that did the music induction based on the brainwave
% Treat music stimuli as a variable, and extract its features.
% Problem:
% a. we never have the same state in our brain
% b. the familiarity of the music changes

\subsection{Emotion-oriented Music Therapy and Recommendation}
Emotion-oriented music therapy helps people achieve a ``delighted'' emotional state (discrete model) or get a higher valence (continuous model). Ramirez et al. introduced a musical neuro-feedback approach based on Schmidt's study to treat depression in elderly people \cite{ramirez2015musical}, aiming to increase their arousal and valence based on the v/a model. The same approach was also applied to palliative care of cancer patients \cite{ramirez2018eeg}. Both studies indicated that music could indeed help patients modulate their emotions to positive arousal and positive valence and thus improve their quality of life. Earlier, Sourina et al. proposed an adaptive music therapy algorithm based on subject-dependent EEG-based emotion recognition, and detected the satisfactoriness of subjects towards the therapy music in real-time \cite{sourina2012real}. They designed the recommender for six discrete emotions in Russell's v/a model, and participants can assign those emotions to the system for getting the corresponding music pieces, similar to the workflow shown on the left side of Figure \ref{compare}. One potential limitation of all those methods is that they all interpreted the user's emotions through a predefined algorithm (e.g., the difference of the alpha power between the left and right scalp to represent the subject's valence level) rather than personalized self-assessment. Another problem is that they acquired feedback from the users after they listened to the music. The predefined algorithm itself was calculated by the collected EEG thus the result could only be an indicator, rather than the benchmark. Instead, self-assessment with the participant's volitional involvement is the most reliable benchmark.

Many other studies have been carried out to recognize music-induced emotions by physiological signals like electrocardiogram (ECG), heart rate, blood pressure, and respiration rate \cite{lin2010eeg, agrafioti2011ecg}. However, none of the prior studies has included people's current emotional state regarding the decision to make music recommendations. People are constantly changing over time, and the interpretation of musical emotion varies among different listeners \cite{soleymani20131000}. Even though people's previous data was collected, music similarity and emotion analysis was conducted \cite{zhu2006integrated}, or the dynamic model of users' preference was built \cite{rafailidis2015modeling}, the event of listening to music is never truly connected with people's emotional state in their equations. We believe that a system should be delivered by the functions of both user and music to benefit users' emotions in an interactive process. 

% =====================================================================
% =====================================================================

\begin{figure}[tbp!]
  \centering
  \includegraphics[width=\columnwidth]{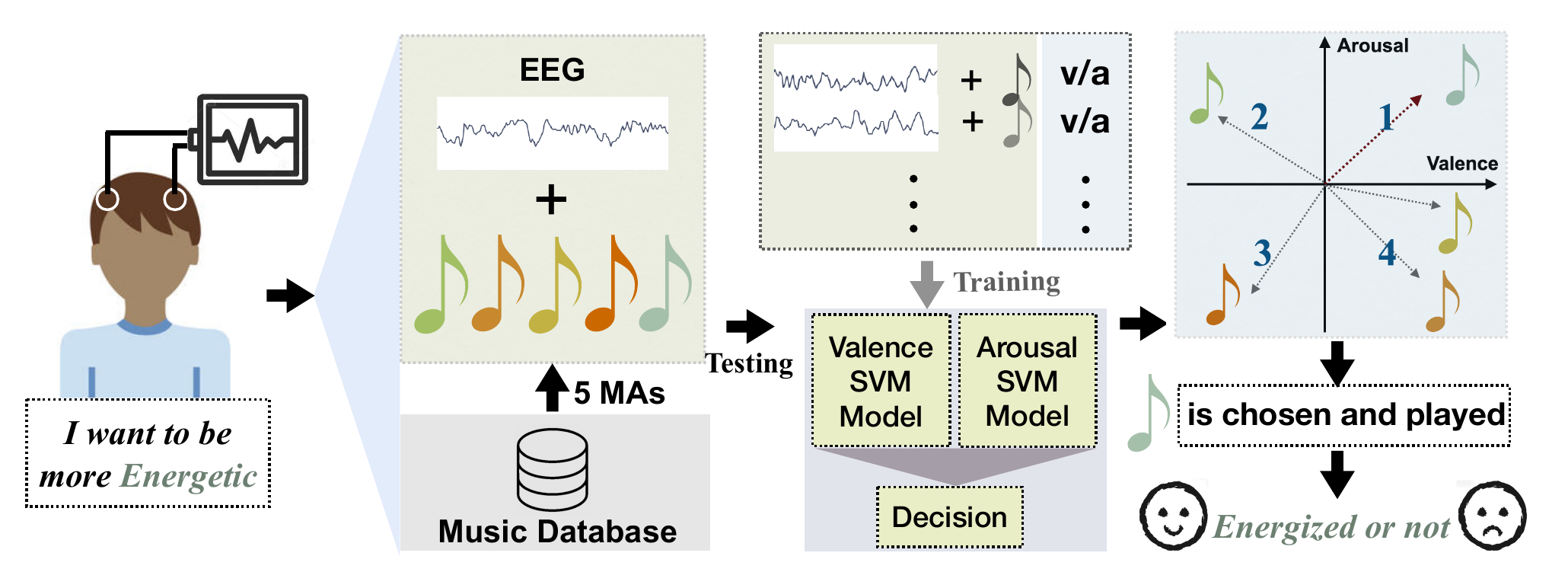}
  \caption{The flow diagram of the training and testing processes in the proposed system. The gray arrow shows the offline process of training, and the black arrows show the online process of testing.}
  \label{flow}
  %\vspace{-15}
\end{figure}

\section{Data Collection}

\subsection{Self-assessment}
% One popular theme in HCI and Ubicomp is the design and development of technologies for mood regulation, such as systems that focus on improving our awareness of our affective states.

Self-assessment/reporting approaches are the most common way to evaluate emotions and serve as the ground truth. Our application demands users to assess their emotion variations induced by a song on Russell's circumplex model, which is marked as the evaluated valence/arousal (v/a) scores as shown in Figure \ref{compare}. Participants were instructed to move two sliding bars ranging from $-5$ to $5$ continuously with scores shown on the bottom after introducing the meaning and difference of ``valence'' and ``arousal.'' The vertical sliding bar is for the arousal score, and the horizontal sliding bar is for the valence score. The window of sliding bars showed up right after the subject listened to each song, and the window also showed up at the beginning of the testing experiment for subjects to designate their desired emotion variations.

% as the system relates to emotion and music, we We choose Russell's circumplex model for users' self-assessment of emotion variation because of its well-known representativeness and reputation it received from studies a wide range of studies \cite{russell1999core, russell1983dimensions, posner2005circumplex}. 

% Even though more dimensional models are discussed \cite{watson1988development, thayer1990biopsychology}. 

\subsection{Music Database}

Music pieces were selected from the ``Emotion in Music Database (1000 songs)'' \cite{soleymani20131000}, which contains 744 songs (excluding redundant pieces). Songs were from eight genres in general: Blues, Electronic, Rock, Classical, Folk, Jazz, Country, and Pop, and were cut into 45-seconds excerpts with a 44,100 Hz sampling rate. The emotion expressed in each song was evaluated on Russell's circumplex model by a minimum of 10 workers found and selected through Amazon Mechanical Turk (MTurk). They both annotated static and continuous emotions expressed in songs. However, the excerpt provided to the user is considered as a variable overall in this paper -- we only predict users' emotional variation after listening to each excerpt, thus we only employ those static annotations as their v/a annotations. To reduce the experiment load and verify the feasibility of the proposed system at the preliminary stage, we provided the first 25 seconds of the 45-seconds excerpts to human subjects, which was long enough to induce listeners' emotions \cite{juslin2001music}. Values of v/a annotations of each music excerpt range from one to nine. We re-scaled them to positive and negative values by minus five, which resulted in four classes/quadrants for coarsely selecting MAs in testing experiments. It should be mentioned that, the v/a annotations here represent the emotion expressed in music, which is decided by the composer and entertainer. In contrast, the v/a scores represent the emotion variations evaluated by users after they listen to a music piece. 

% Music is a dynamically changing stimulus thus users' emotion would be changed variously while listening to it. So t

% the integrity of the song where the excerpt comes from and static annotations are utilized. Based on the ideThe first 15 seconds of continuous annotations were excluded due to the instability of the annotations at the start of the clips. We used the first 25 seconds of music excerpts in the experiment, therefore, the static v/a annotations were chosen to represent the music emotion. 

\subsection{Participants \& Procedure}
A total of five participants (4 males; aged 22 to 31 years) were recruited to participate in the experiment, and there is no clinical (including psychological and mental) problem reported by those participants. Participants were labeled as \textit{s01} to \textit{s05}, respectively, and the model was built for each participant individually. The training and testing processes were different: the former one was to collect data and build the prediction model offline, and the testing process was an online process that needs to collect and analyze the data in real-time to recommend music. To testify the robustness of the system towards user's varying emotional states, we split the experiment into 13 days: six days of training experiments, two days of testing experiments, and the other five days of the real-life scenario evaluations that were modified from testing experiments. In addition, 54 people, including the five participants, were invited to participate in an online questionnaire study. They were asked to respond to 10 questions according to their own experiences independently and anonymously. The responses were used to investigate the public's demand and acceptance of emotion regulation with an intelligent music recommendation system and wearable EEG device. All of the experiments were approved by the Internal Review Board (IRB) at [University is hidden for double-blinded review]. 

% The results from these five participants are going to show the usability and functionality of our EEG-based music recommendation system with four classes of emotional states. 

Participants were asked to understand the arousal and valence emotions respectively, with the help of the discrete emotions labeled in Russell's model as shown in Figure \ref{Russell} before the experiment. And there is no restriction regarding their normal music listening experiences or the emotional state before an experiment. As a factor of music effects, music volume was adjusted by participants to their habitual level before the experiments.

To build the prediction model offline, songs in the training process were chosen randomly from each quadrant based on their v/a annotations. A trial of a training experiment had three sequential parts: EEG data collection (20s), music playing (25s), and users' self-assessment (15s). EEG data collected and a music piece played in each trial formed an observation in training data, and users' self-assessment serves as the ground truth. Considering the comfortable wearing time of the Emotiv EPOC+ device is about half an hour, we designed twelve trials (each trial lasts for one minute) for each experiment, in which twelve songs were from four quadrants evenly, and two experiments for each day. Plus five-minute EEG device setup and three-minutes rest between two experiments, in total 32 minutes were required for one subject each day. EEG data were collected at the Resting State with Open Eyes (REO), and the next trial started right after the self-assessment. 
% The model is unable to work for song recommendation while collecting training data as shown on the upper-right of Figure \ref{compare},  

% Three songs are , so that twelve songs corresponding to twelve trials are selected averagely for four classes. 
% Thus each trial contains the information of one pair of EEG-song. After that, participants were requested to evaluate their emotion variations (i.e., v/a scores) as a result of listening to the song. After each experiment, the song came from one class but changed the participant's emotion to another class was indexed into  the designated v/a class in the training process was indexed into the candidates. 

% Based on the positive/negative value of the v/a scores, we separate users' emotional states into four classes. 

After training the emotion variation prediction model, participants took part in the testing experiments, as shown with the `black' arrows in Figure \ref{flow}. Testing was a real-time process and there were 22 trials for each testing experiment. Each trial had seven steps as shown on the lower-right of Figure \ref{compare}: (1) the participant designated the desired emotion variation (designated v/a scores); (2) five music alternatives (MAs) were randomly selected from the same class as the designated v/a scores based on their v/a annotations; (3) collected EEG data; (4) extracted EEG features and concatenated with the features of each music alternative; (5) concatenated feature vectors are input into the prediction model sequentially and returned five results; (6) the song of which the predicted class of v/a scores was located in the user's designated class was played; (7) participants reported the emotion variations based on the presented song.

% and music features were extracted offline. with the online feature extraction of participant's current EEG data and prediction of emotion variation. T

% The time of each trial in testing experiments depends on the computation efficiency of the real-time processing, as well as whether there is at least one song 
It is possible that no song in MAs matches the participant's desired emotion variation. If so, the system will go back to step (2) and repeat the following steps until a new song matches. The maximum iteration was set to five to avoid the system delay, and a song from the fifth iteration will be randomly selected if still no song matched. In the case when several songs matched, the system randomly selected one. Thus, the time of each trial in testing experiments was unfixed. And the proper time length of EEG collection between playing music pieces will be decided by referring to the classification accuracy of training experiments with different time windows of EEG data.  

% And we train the prediction model by make use of the different length of EEG data in training experiments accuracy  based on different time-windows of EEG data in `Classification Model' Section. 

% Since the testing experiment is a real-time system that designed to be used in people's daily life, the interruption between songs should be as insignificant as possible.  For our system aims to be an application that people can use in their daily life, experiments of a real-life scenario were processed to explore questions that will appear in daily usage after testing experiments. 
\begin{figure}[!tbp] 
% \centering
    \begin{subfigure}
    \centering
        \includegraphics[width=.31\linewidth,height=.24\linewidth]{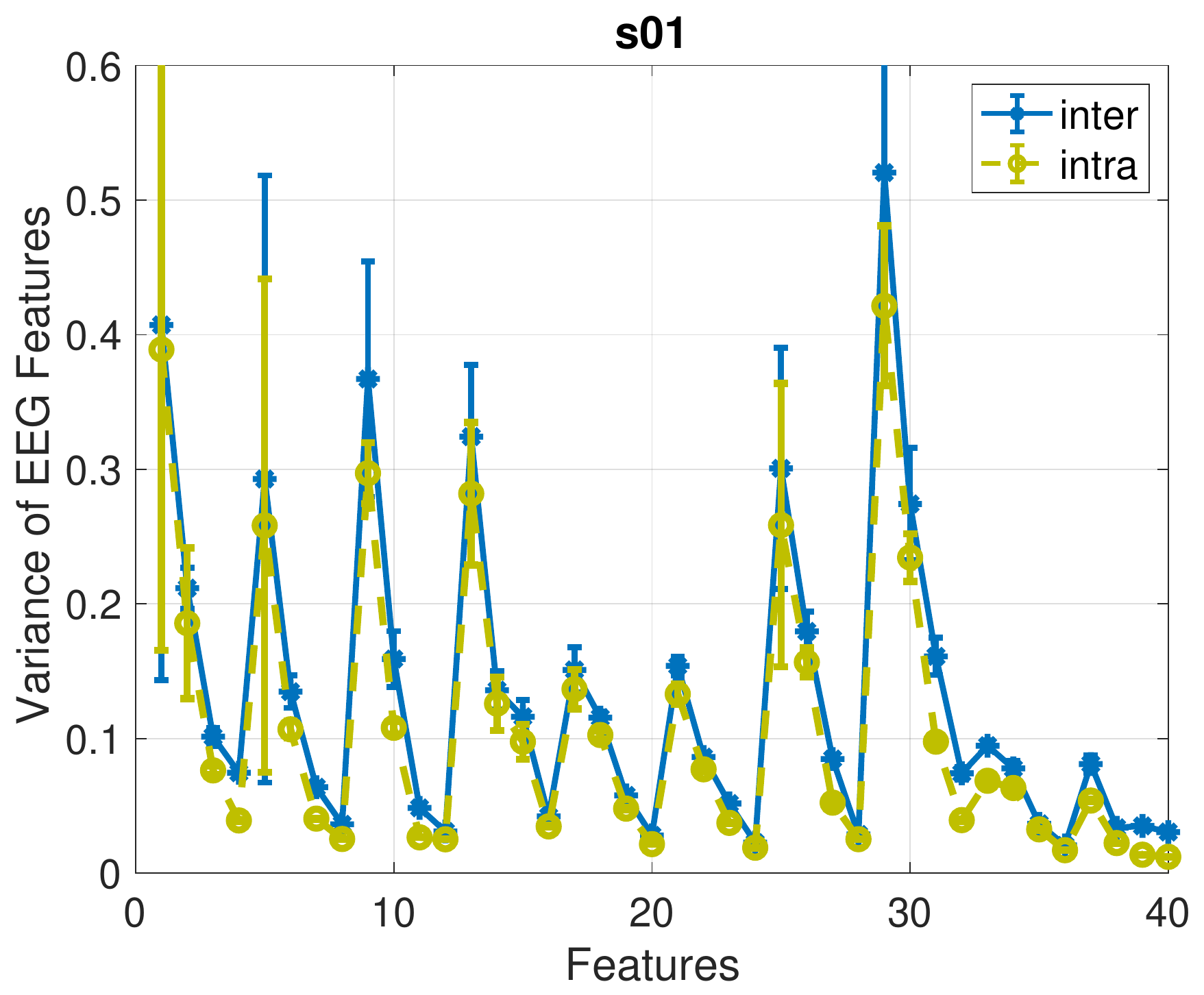}
        % \caption{s01}
    \end{subfigure}%
    ~ 
    \begin{subfigure}
    \centering
        \includegraphics[width=.31\linewidth,height=.24\linewidth]{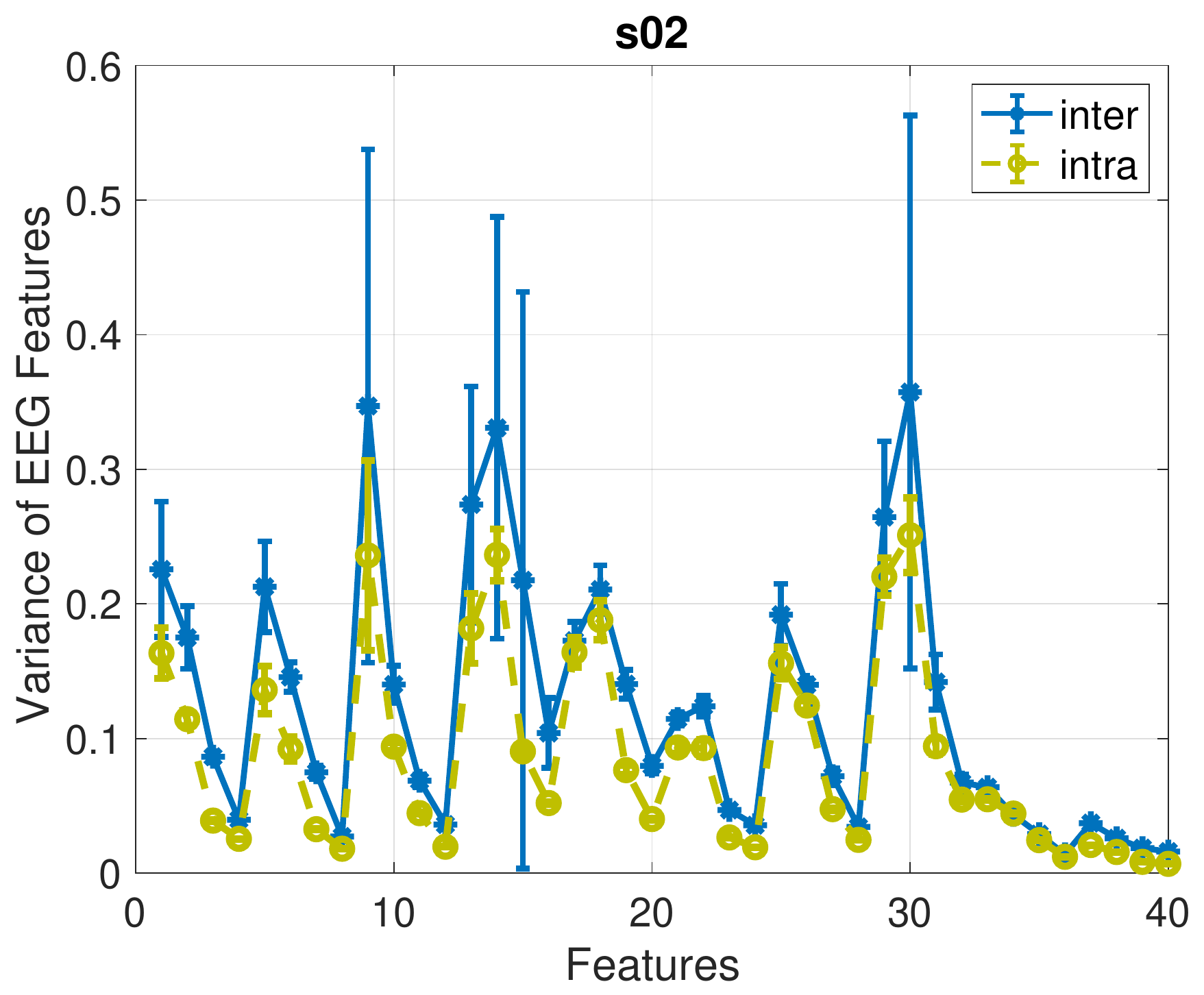}
        % \caption{Lorem ipsum, lorem ipsum,Lorem ipsum, lorem ipsum,Lorem ipsum}
    \end{subfigure}
     ~ 
    \begin{subfigure}
    \centering
        \includegraphics[width=.31\linewidth,height=.24\linewidth]{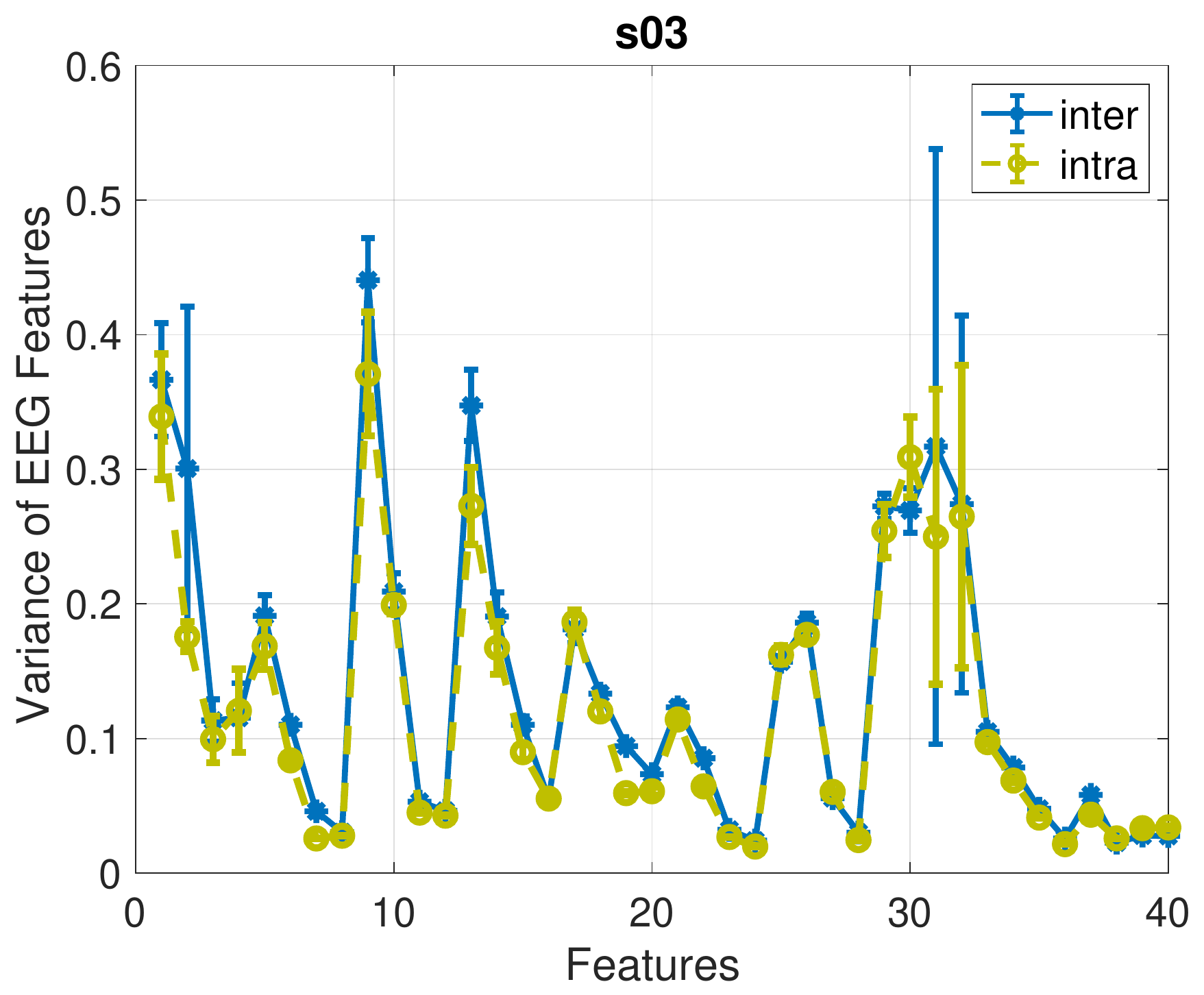}
        % \caption{Lorem ipsum, lorem ipsum,Lorem ipsum, lorem ipsum,Lorem ipsum}
    \end{subfigure}
     ~ 
     \newline
    \begin{subfigure}
    \centering
        \includegraphics[width=.31\linewidth,height=.24\linewidth]{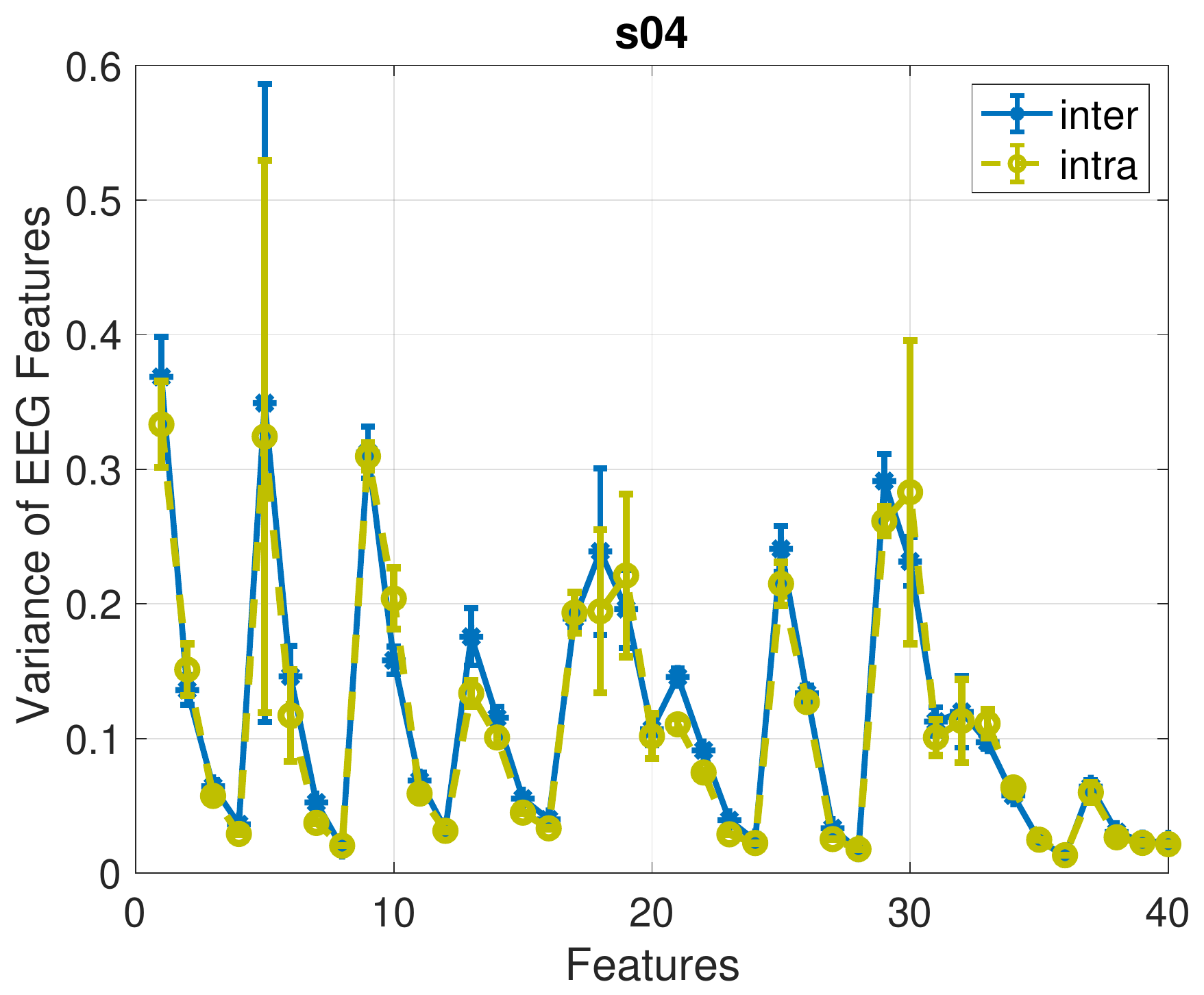}
        % \caption{}
    \end{subfigure}
     ~ 
    \begin{subfigure}
    \centering
        \includegraphics[width=.31\linewidth,height=.24\linewidth]{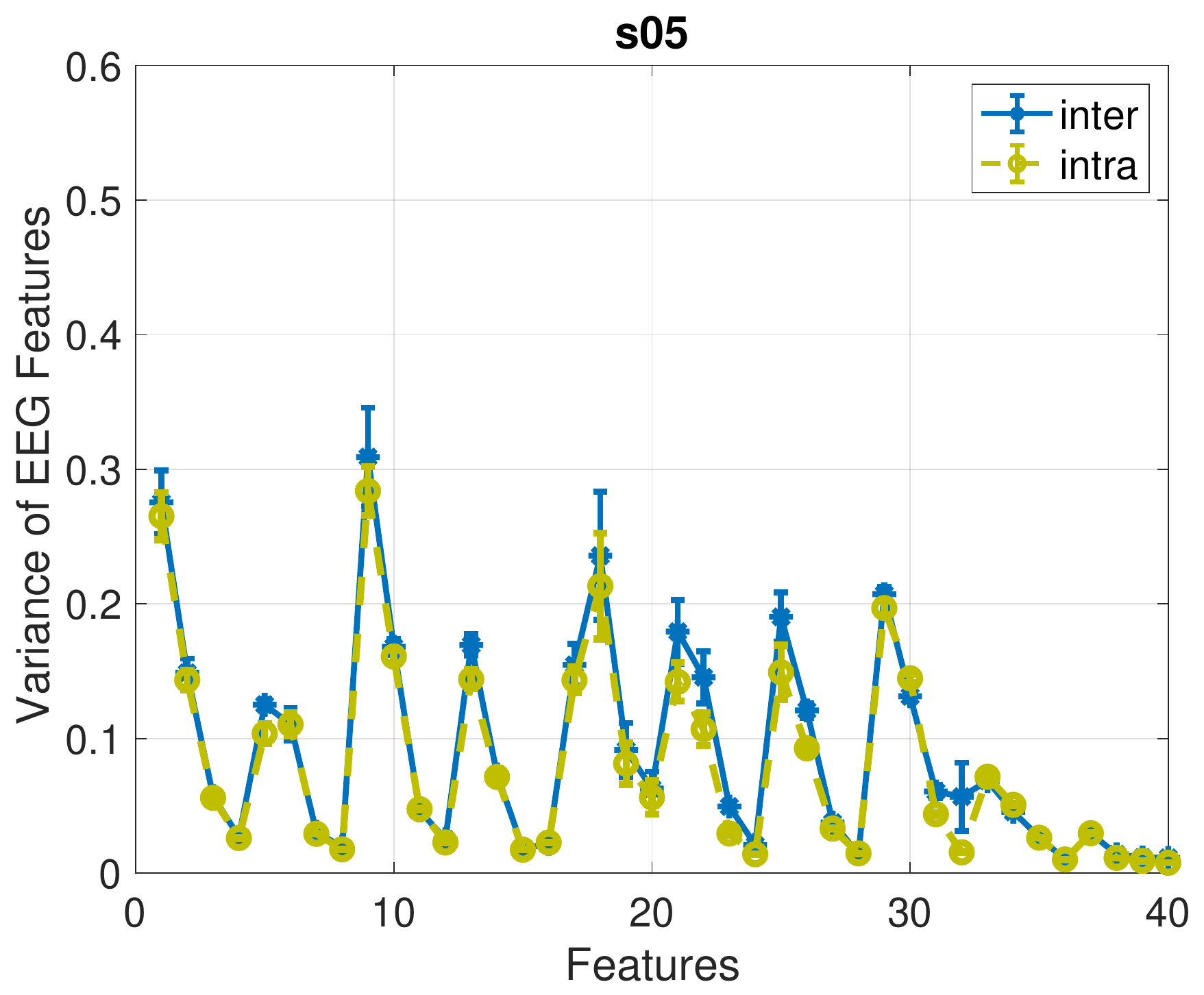}
        % \caption{}
    \end{subfigure}
    \caption{Comparison of mean-variances of EEG features between intra-experiments and inter-experiments for five participants. The x-axis is the extracted EEG feature and there are 40 in total.  The y-axis is the mean and standard deviation of feature variances. The intra-variances are calculated from trials in one experiment, and the inter-variances are calculated from trials in two continuous experiments.}
    \label{fig:var}
\end{figure}
Testing experiments were modified to accommodate the requirements of a more usable system in the real-life usage scenario. In this scenario, users only designated the desired emotion variation once, and EEG data were collected once at the beginning of the system usage to reduce users' active interaction efforts. And the ``current'' information from users was used to select a list of songs instead of one song. We assumed that the desired emotion variation would sustain, and one-time EEG data would be representative of users' current state within one experiment. To verify this hypothesis, we calculated and compared the variance of EEG features extracted from the training sessions within one experiment (intra) and two experiments (inter). The inter-test was combined with the tail half and head half trials of two continuous experiments to guarantee the same number of trials as the intra-test. We drew the mean-variance of intra- and inter-experiments for each participant, as shown in Figure \ref{fig:var}. It is shown that, on average, for five participants, $89.5\%$ of $40$ EEG features in intra-experiments have lower mean variance than the ones in inter-experiments. We thus assume that, with reasonable variance, EEG information is representative of users' ``current states'' and one-time EEG collection would work properly within twelve pieces of music (one experiment) for real-life scenarios.

\section{Methods}

\subsection{Music Feature Extraction}
We extracted and selected features of each music piece by using Librosa \cite{mcfee2015librosa}. Each song was a one-channel MP3 file with a sampling rate of 44,100 Hz. Both time-domain and frequency-domain features were included, as shown in Table~\ref{tab:mus_fv}. Chromagram is a mid-level feature, which closely relates to ``twelve-pitch classes''. Tempogram represents the local auto-correlation of the onset strength envelope, and thus holds an important position for its rhythm information. In the case of roll-off frequency, different percentages refer to the roll percent of the energy of the spectrum in that particular frame. We chose four different values of the roll-off percentage to approximate the maximum and minimum frequencies in different ranges. The music feature matrix was normalized before further processing. And the feature selection and reduction were processed through the following steps: (1) removing constant columns which have $100\%$ same values; (2) removing quasi-constant features with threshold $90\%$; (3) removing columns with the Pearson correlation greater than 0.92. The number of extracted features is 1,665, and the number of selected features after the above steps is reduced to 174. 

\subsection{EEG Feature Extraction}
Emotiv EPOC+ provided 14 channels of EEG data (AF3, F7, F3, FC5, T7, P7, O1, O2, P8, T8, FC6, F4, F8, and AF4) with 128 Hz sampling rate. Raw EEG data were first high-pass filtered by 8 Hz and then low-pass filtered by 49 Hz through EEGLAB \cite{delorme2004eeglab}, as alpha (8–13 Hz), beta (14–30 Hz), and gamma (31–50 Hz) power bands are proven to be relevant with emotion \cite{kim2013review}. Noises of eye blinks (around 3 Hz) were rejected by the high-pass filter, and high-frequency noises like eye movements, muscle activity, and environmental noise were rejected by the low-pass filter. The artifact was rejected by implementing ``Reject continuous data'' on EEGLAB, which used the spectrum thresholding algorithm to detect artifacts. The artifact channels were detected by implementing automatic channel rejection, which used the joint probability of the recorded electrode with ``kurtosis'' and returned the measured value for each channel, and we rejected channels with measure values higher than 20.
\begin{table}[!tbp]
\centering
  \caption{Extracted music features in the time- and frequency-domains}
  \label{tab:mus_fv}
  \begin{tabular}{l r}
    % \toprule\begin{figure}[tbp!] 
% \centering
    Features & Number\\
    \midrule
    Zero-Crossing Rate & 1\\
    Bandwidth order 2,3,4 & 3 \\
    Mel-Scaled Spectrogram & 128\\
    MFCC &13 \\
    Chromagram from STFT & 12\\ 
    Root-Mean-Square & 1\\
    Spectral Centroid & 1\\
    Spectral Contrast & 7\\
    Spectral Flatness & 1\\
    Spectral Roll-off 5,10, 85, 95 ($\%$) & 4\\
    Tempogram &384\\ 
  \bottomrule
\end{tabular}
\end{table}
% \subsubsection{Feature Extraction and Selection}
% Users' current EEG information was acquired before listening to music. Considering that the EEG data collection would make the interruption too significant in the real-time system, we need to determine the a proper length of EEG data that used for emotion variation prediction. To do so,

Before feature extraction, we defined three time windows in order to determine the proper time length for EEG data collection in online testing experiments. To accommodate the lowest frequency of interest, the half cycle should fit comfortably within the window \cite{budzynski2009introduction}. Normally 0.5 Hz is taken as the low end, which mandates a time window of at least one second. And we double the low end to 2 seconds as some data with artifacts would be rejected by pre-processing. The other two window lengths are 5 and 10 seconds, respectively.

% we evaluate the accuracy of prediction model with three different time windows on training data
 
% Information of EEG was using Fourier transform and sufficient samples were converted to its constituent frequencies.  To provide sufficient but not too lengthy EEG samples, we extracted features from three time-windows of $\{10, 5, 2\}$ seconds and evaluated the validation accuracy, then decided one time-window for the testing experiments. The frequencies in alpha (8–13), beta (14–30) and gamma (31–50) bands were representative for EEG frequency from 8Hz to 48 Hz. However, the range of beta band was relative coarse so we

Features were extracted based on three window lengths of EEG data separately. Fast Fourier Transform (FFT) was employed with 3, 2, and 1 point moving-average corresponding to the time windows of $\{10, 5, 2\}$. Then we re-grouped the frequencies into four bands: alpha (8-12 Hz), beta-1 (12-16 Hz), beta-2 (16-32 Hz), and gamma (32-48 Hz) bands. The first set of features are power differences of each band between the right and left channels: AF4/AF3, F8/F7, F4/F3, and FC6/FC5, because differences between the right and the left frontal lobe were proved to be related with valence in many research \cite{henriques1991left, davidson1992emotion, allen2004issues}. The second set of features is the band power of electrodes placed on the temporal lobe: T7, T8, and the neighboring sensors: P7, P8, for the reason that the temporal lobe was directly related to the region where the sound is processed \cite{gloor1997temporal}. The last set of features is the mean and standard deviation (std) of the power of all channels. Thus the total number of EEG features is $40$ $(10 \times 4)$, and they were normalized feature-wise before being combined with music features.

% Before feature selection, we need to choose the proper period of EEG data collection among $\{10, 5, 2\}$ seconds. And we choose to refer to the validation accuracy of classification model that based on those three lengths of features. Data from O1 and O2 are only included in the last set because they are located on the occipital lobe and have the least relation with emotion.  that is, the song listened to after the EEG collection in the same trial

% by a Python library named mlxtend

Sequential Backward Selector (SBS) was utilized to decrease the dimension of music features before being concatenated with EEG features in order to balance the amounts of music and EEG features. SBS is a greedy search algorithm removing the feature one by one to maximize the performance until reaching the number of features that were designated. Three time windows $\{10, 5, 2\}$ of EEG features were concatenated with selected features of the music piece in the same trial separately. Features for the arousal model and valence model were selected by SBS independently from concatenated features for each participant. 

% And we did feature extraction and selection steps for all three time-windows $\{10, 5, 2\}$ of EEG. The remaining number of music features is 50 

% for choosing the right period of EEG collection in next step

\subsection{Classification Approach}
Before testing, training data was used to build the classification model and decide the optimal time length for EEG data collection. As we discussed, predicting continuous emotion variation values is impractical and infeasible. For example, $[-0.1, -0.1]$ and $[0.1, 0.1]$ are two v/a scores, they are close in a regression model but distinct in terms of the direction of emotion variation. The former one would mean that ``the music makes me kind of blue'' (3rd class), and the latter one means that ``the music makes me slightly delighted'' (1st class). Besides, neither the value of the designated v/a score nor that of the evaluated v/a score could the participant be firmly certain about. Instead, participants are more confident about the direction or range of valence and arousal changes. 

We chose the binary classification for valence and arousal models because the boundary of emotions on the continuous model is undefined. The problem appears for two close v/a scores when another hard boundary is defined other than x- and y-axes. For example, if a line separated two relatively far emotions ``Aroused'' and ``Happy'' in the first quadrant, then it would also distinguish two similar emotions near the line into two different classes. Support Vector Machine (SVM) was employed due to its great generalization capability, computational efficiency, and proven performance for EEG-based emotion classification \cite{bazgir2018emotion}. The formation of the arousal model is shown in Equation \ref{equ:1}:

% Features of valence and arousal classification models were selected individually with same methods introduced in ``EEG Feature Extraction'' Section. And the class of the final emotion variation was decided by two branches resulting in one of the four quadrants as shown in Figure \ref{flow}.

% Reduced feature matrices from $\{10, 5, 2\}$ seconds' of EEG data were input into the emotion variation prediction model individually. Even though users assess their emotion variation on Russell's v-a model,

% our prediction model will based on the extracted emotion types (four classes locate on four quadrants) Classification models were selected instead of regression models for the consideration that the positive or negative effects of music for users were more representative than the value itself. 

% Additionally, we wanted to show the outcome from a relatively simple model and concentrate on the relation between EEG-music and emotion variation, which subjected to such a preliminary study using EEG-music pair for the recommendation. 
\begin{equation}
   y_a = C_a' * [E_a, M_a]
    \label{equ:1}
\end{equation}
where $y_a$ represents the binary class of arousal, and $C_a$ is the parameter vector of the arousal model. Selected feature vectors of EEG and music for the arousal model are represented as $E_a$ and $M_a$ and they are concatenated together to learn the emotion variations. The valence model is expressed with the same equation. To manage the unbalanced observations collected for training, the regularization parameter was adjusted by the ratio of their binary labels. It is worth pointing out that, since the data collection of the testing process was separated from the training process, all of the observations collected in training experiments were only used to train the model.

% And the measure Model can be overfitted by collected EEG and selected musics The regularization parameter C is chosen by the 
% EEG is different from any other and we can never see it completely, thus, to avoid overfitting the classification model, it is beneficial to evaluate  generally to adjust the regularization parameter.  

\subsection{Emotional Instability}
In addition to predicting the direction of valence and arousal changes, a secondary study of this research is related to participants' emotional instability. There were $25\% - 35\%$ of songs that reappeared two to five times for each participant in different training experiments. And their varying feelings towards these songs were employed to quantify users' emotional instability. People listen to different songs at different times in real life, thus we set a range for the repeated songs and the repeating times when randomly selecting songs in training experiments. And we assume the same number of repeated songs and repeating times for all participants will mislead us with highly biased results and can't draw a general conclusion of whether the emotion instability scores are indicative in future usage.

To quantitatively represent and assess the emotional instability, the number of transitions of arousal/valence scores was counted. The arousal or valence scores of a song have two states, $i=0$ (negative) or $i=1$ (positive). We propose that the frequency of transitions of 0 to 1 and 1 to 0 towards the same song reveals the user's emotional instability. For example, the arousal score vectors $[0,0,0,1,1]$ and $[0,1,0,1,0]$ of a song have the same probability of 0 and 1, while the first score vector tends to result from a change in the subject's taste after listening to this song for three times, which is less related to the emotional instability compared with the second subject. We define the frequency of transitions as t-score and it is calculated following the Equation \ref{equ:2}:
\begin{equation}
    t=\frac{1}{M}\sum_{m=1}^{M}(\frac{1}{N}\sum_{n=2}^{N}|(s_n - s_{n-1})|)
    \label{equ:2}
\end{equation}
where the score vector $s$ has the binary pattern, thus the summation of the absolute difference of two neighboring elements is the number of transitions. $N$ is the number of times the song with $id=m$ is listened to by a subject. $M$ is the total number of repeated songs. By implementing the equation, the results of the two vectors mentioned before are $0.25$ and $1.0$, respectively, which is consistent with our assumption that the latter subject shows a higher level of emotional instability than the former subject. 

% After calculating the arousal and valence transition values of participants, we employed the Big Five Personality Test to see whether the transition value could be a valid indicator of emotion instability. 

To verify the proposed emotional instability function from a psychological perspective, we refer to the Big Five Personality Test \cite{Goldberg1992} based on participants' self-reporting and psychological test on Open-Source Psychometrics Project \cite{OSPP}. Big Five Personality Test is based on the five-factor model of personality \cite{mccrae1999five}, including openness to experience, conscientiousness, extraversion, agreeableness, and neuroticism. Among all five factors, neuroticism is referred to in our study because it describes emotional stability. Participants were required to finish the Big Five Personality Test within ten minutes and their respective scores of neuroticism were collected. The score is defined in a way as what percent of other people who have ever taken this test and performed worse than you; that is, a higher score indicates a more stable emotional state. We compared the t-score and the neuroticism score among all participants to verify their correlation level and thus prove that our study can be used as an indicator of an individual's emotional instability. To correspond to the instability scales calculated by Equation~\ref{equ:2}, the neuroticism score is rewritten as $(1-score/100)$.

% or the proportion of music pieces that matched the user's demands,

\subsection{Evaluations}
Training accuracy of the valence and arousal models are shown by cross-validation separately. To validate the interactive performance of the system between the user and recommended music more directly, testing accuracy is decided by both valence and arousal models: the match rate between the designated v/a score and the evaluated v/a score. In addition, to explore the feasibility of using fewer EEG electrodes, we selected electrodes on the temporal lobe region: T7 and T8, and evaluated the training accuracy solely based on these two sensors. Lastly, we evaluated the performance of the real-life scenario, which was carried out two months after the original testing experiments for five continuous days to meet participants' schedules. The same five participants were recruited and their models were trained by all the data collected before. Meanwhile, participants listened to new \& old songs (which were listened to in the training and testing experiments or not) from the same database. The accuracy was evaluated using the match rate as the testing experiments, and the emotional instability score was updated with new data.

\section{Results}
% The first three rows are the statistics from training experiments and the last three rows are from testing experiments. \textit{\#Train\_song} and \textit{\#Test\_song} are the numbers of songs participants listened to in training and testing experiments. \textit{\#Uniq\_song} is the number of unique songs listened to in training experiments. \textit{Match\_rate} represents the percentage of music pieces, of which the v/a annotations match the user's emotion variation. \textit{\#New\_song} is the number of songs never listened to in the training experiment. \textit{Test\_accu} is the test accuracy of our recommendation system.
\subsection{Data Collection \& Feature Selection}
\begin{figure}[tbp!]
  \centering
  \includegraphics[width=0.6\columnwidth]{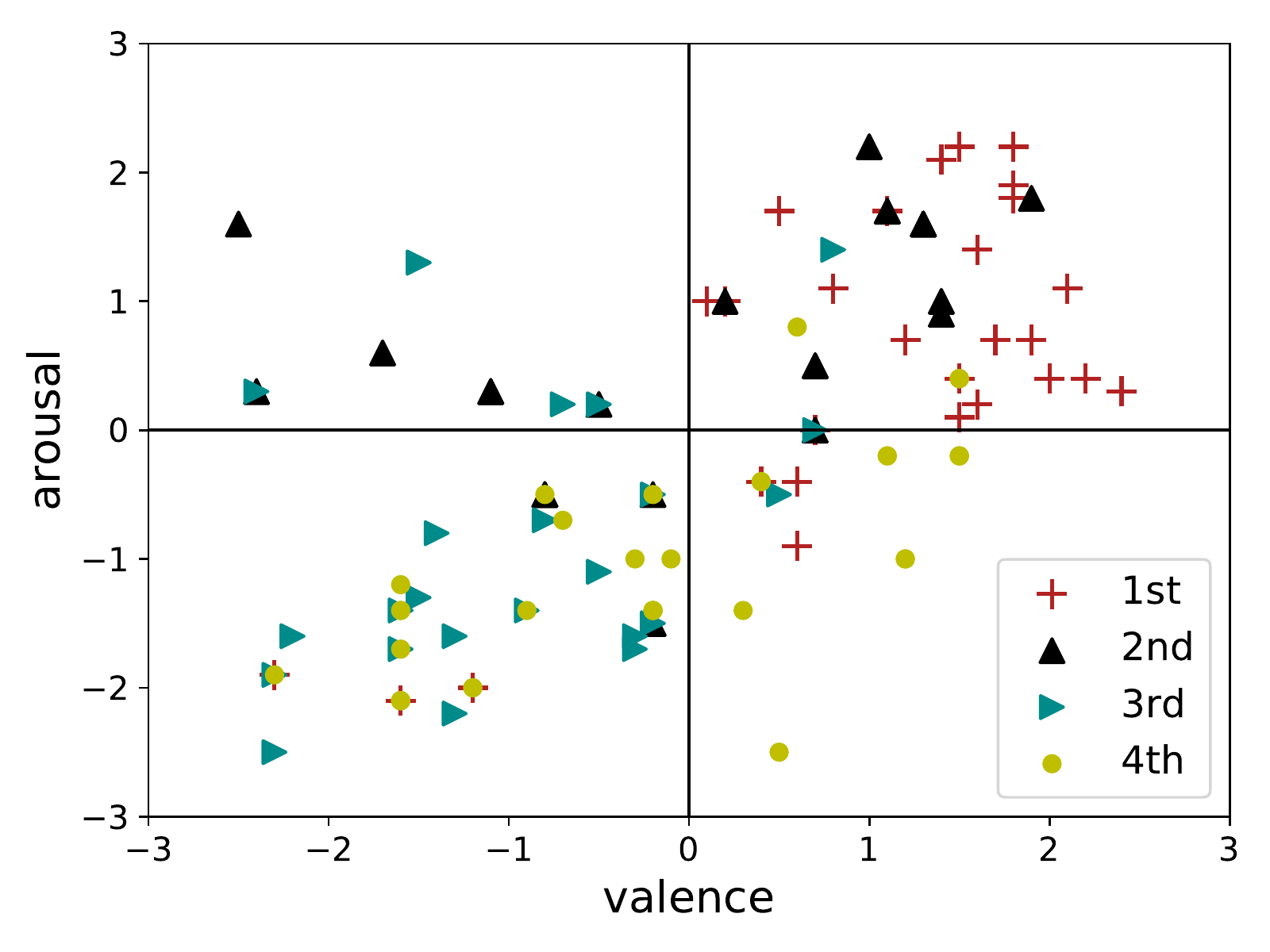}
  \caption{Music excerpts listened by \textit{s01} in training experiments located on Russell's valence-arousal model by their v/a annotations. They are marked by \textit{s01}'s evaluated v/a scores. The different markers on the same location represent the participant's different v/a scores for the same song. }
  \label{dist}
\end{figure}

Trials in training and testing experiments with rejected channels are excluded, and the remaining number of trials are shown in Table~\ref{tab:accu} as (\textit{\#Train\_song} and \textit{\#Test\_song}). Each trial represents an observation that contains the collected EEG data and the song presented to participants. The first three rows of Table~\ref{tab:accu} are the statistics from training experiments. \textit{\#Uniq\_song} shows the number of songs without any repetition. \textit{Match\_rate} refers to the percentage of music pieces that have the same v/a annotations (the emotion expressed in the song) with users' evaluated v/a scores (the emotion evoked by the song). The mean match rate of five subjects was $0.4938 \pm 0.0504$, which was close to $0.5$, we arguably conclude that recommending music based on only music v/a annotations for user's emotion variation resembles a random occurrence. To show the match rate visually, we plotted the training data of \textit{s01} in Figure~\ref{dist}, in which each data point represents a song that is located by its v/a annotations, and the marker represents the class of evaluated v/a scores. It shows that the participant's emotional states could be varied to all four classes by songs from the same quadrant, and the song with overlapped markers means that it can vary the participant's emotions oppositely in different experiments.

The last three rows of Table~\ref{tab:accu} are the statistics from testing experiments. \textit{\#New\_song} is the number of songs that have never been listened to in the training experiment (new to the prediction model). The match rate calculated in testing experiments is used to evaluate the system performance and called (\textit{\#Test\_accu} to distinguish it from the match rate of training experiments.

The number of EEG features and music features are 40 and 178 after feature extraction. Fifty music features were eventually selected by SBS based on v/a annotations before being concatenated with EEG features. Considering the limited number of observations we obtained, the final number of features was set to 25 by implementing SBS on the concatenated features for both valence and arousal models. The remaining EEG and music features for the valence and arousal models of each participant are different. The remaining number of EEG features in arousal model is $[6, 8, 10, 13, 7]$, and in valence model is $[8, 10, 11, 5, 14]$ for participants from $s01$ to $s05$. And the first 25 most significant features would change if we re-train the model with new data.

% Besides computing the classification accuracy by training data, we also reveal the low match rate of music emotion (v/a annotation) and users' emotion variation (v/a score) in training experiments to demonstrate the improper music recommendation strategy that depends on music emotion only. The result is shown in Table \ref{tab:accu}. The mean match rate of five subjects was $0.4938 \pm 0.0504$, which was close to $0.5$. So we arguably conceive that recommending music based on music v/a annotation resembles recommending music randomly (match/not match). To show the match rate visually, we plotted the songs listened by \textit{s01} in Figure~\ref{dist}. For each song (i.e., a data point in the figure), the location is determined by its v/a annotation, and the marker represents the class of \textit{s01}'s evaluated v/a score. It can be observed that, songs in each quadrant can vary the participant's emotion to any of the four classes, and the song with overlapped markers varies the participant's emotion into different classes in different experiments.

% The last row of Table~\ref{tab:accu} is the match rate of the designated v/a score and the evaluated one in testing experiments, and we call it \textit{Test\_accu}. It validates our strategy of including users' current EEG information for music selection. 

% and labels are participants' evaluated v/a score For later

\begin{table}[tbp!]
  \centering
   \caption{Training and testing statistics of experiments}
  \begin{tabular}{c c c c c c}
    % \multicolumn{6}{c}{\small{\textbf{Statistics of Experiments}}} \\
    \toprule
    % \cmidrule(r){1-6}
    % \hline 
    {\small\textit{}}
    & {\small\textit{s01}}
    & {\small \textit{s02}}
    & {\small \textit{s03}}
    & {\small \textit{s04}}
    & {\small \textit{s05}}\\
    % \hline \hline
    \midrule
    {\small \textit{\#Train\_song}} & 115 & 129 & 135 & 131 & 131\\
    {\small \textit{\#Uniq\_song}} & 72 & 85 & 107 & 79 & 88\\
    {\small \textit{Match\_rate}} & 0.522 & 0.512 & 0.504 & 0.405 & 0.526\\
    {\small \textit{\#Test\_song}} & 44 & 40 & 40 & 41 & 43\\
    {\small \textit{\#New\_song}} & 33 & 31 & 35 & 31 & 32 \\
    {\small \textit{Test\_accu}} & 0.867 & 0.850 & 0.850 & 0.976 & 0.884\\
    % {\small \textit{Old\_song}} & 11 & 9 & 5 & 10 & \\
    % {\small \textit{Unsat\_old}} & 0 & 1 & 0 & 0 & \\
    % {\small \textit{a\_acc}} & 0.956 & 0.950 & 0.949 & 0.976 & 0.932 \\
    % {\small \textit{v\_acc}} & 0.911 & 0.900 & 0.897 & 1.000 & 0.955 \\
    % \hline
    \bottomrule
  \end{tabular}
  \label{tab:accu}
\end{table}

\begin{figure}[tbp!]
  \centering
  \includegraphics[width=0.6\columnwidth]{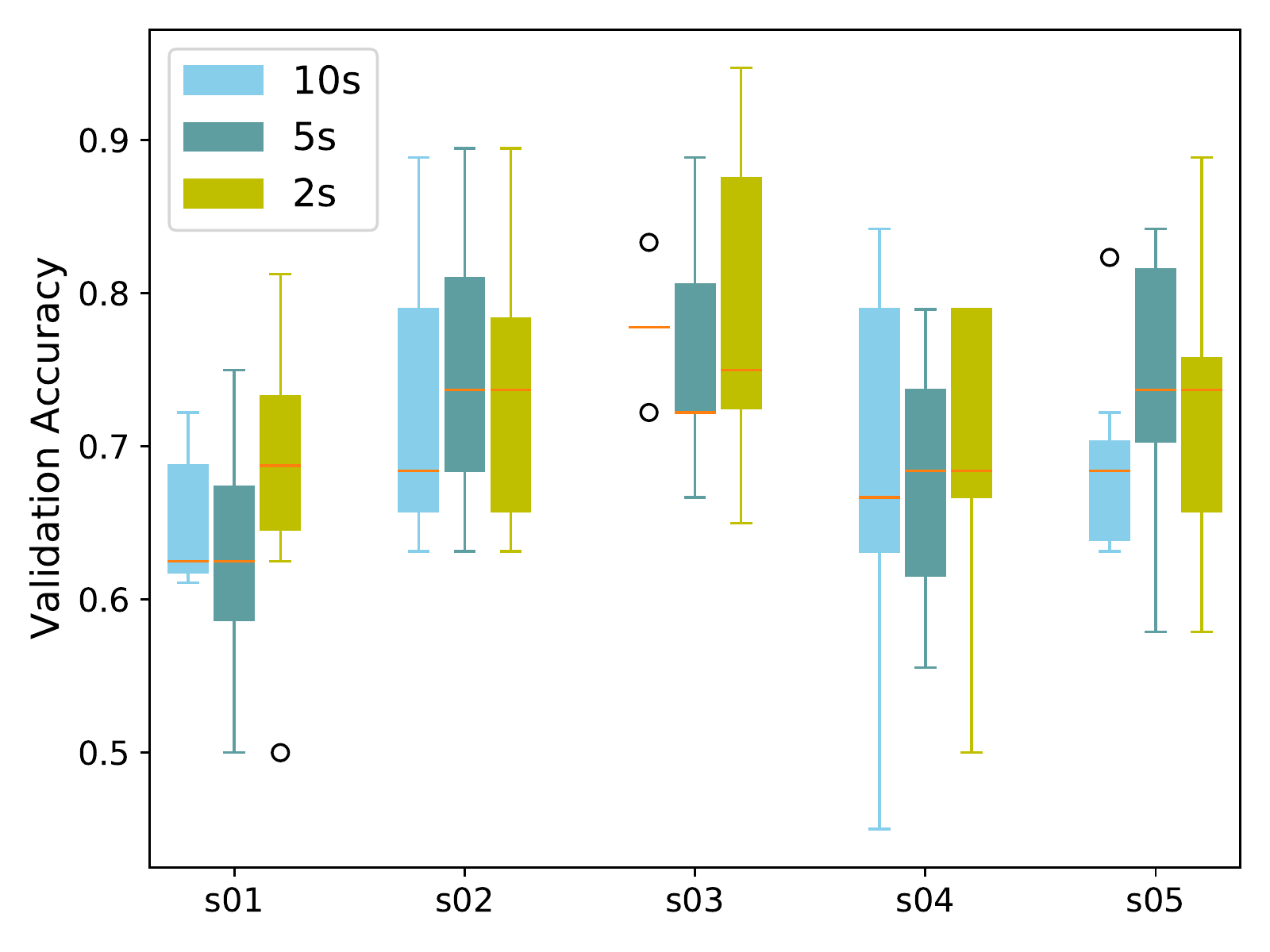}
  \caption{The validation accuracy of arousal model with three time-windows evaluated by 7-fold cross-validation.}
  \label{7cv}
\end{figure}

\subsection{Classification Model}

The arousal model was trained by $\{10, 5, 2\}$ seconds of EEG data separately to select the proper time length for testing experiments. The result of the 7-fold cross-validation for the arousal model is shown in Figure \ref{7cv}. The F-value and P-value for the classification accuracy of three different lengths are $0.321$ and $0.726$, which cannot reject the null hypothesis that the distributions under these three conditions are normal, thus there is no time length better or worse in this experiment. Even though the longer, the more EEG information is contained, the window lengths of 5s and 10s don't outperform 2s for representing users' current emotional state. Thus, we chose the shortest time length for testing experiments. The same conclusion was obtained for the valence model. 

Besides, we calculated the classification accuracy of training experiments with only two channels (T7 and T8) and compared it with 14 channels to verify the possibility of using only a small number of electrodes. In addition to the 8 EEG features from T7 and T8 channels, we selected the 11 most significant features of music by SBS and concatenated them with EEG features. The results are shown in Table~\ref{tab:channel}. With a slight decrease in classification accuracy in both valence and arousal models, the system retains a reasonable level of accuracy with only 2 EEG channels, even though they contain 7 times less information than the original 14 channels.

\begin{table}[tbp!]
\centering
  \caption{Comparison of validation accuracy between 14-channel (`FS') and 2-channel (`T78') EEG data}
  \label{tab:channel}
  \begin{tabular*}{245pt}{@{\extracolsep{\fill}}c c c|c c}
    \toprule
    % \cmidrule(r){1-5}
    % \hline
    {\small\textit{}}
    & \multicolumn{2}{c}{{\small\textit{Arousal}}} 
    & \multicolumn{2}{c}{{\small\textit{Valence}}}\\
    % & {\small\textit{Arousal}}
    % % \cmidrule(r){4-5}
    % & {\small\textit{Valence}}\\
    \hline \hline
    % \midrule
    {\small\textit{}}
    & {\small\textit{FS}}
    & {\small \textit{T78}}
    & {\small \textit{FS}}
    & {\small \textit{T78}}\\
    \hline 
    % \midrule
    {\small\textit{s01}} & 0.68 $\pm$ 0.19 & 0.64 $\pm$ 0.19 & 0.64 $\pm$ 0.22 & 0.63 $\pm$ 0.22\\
    {\small\textit{s02}} & 0.74 $\pm$ 0.18 & 0.73 $\pm$ 0.22 & 0.67 $\pm$ 0.11 & 0.62 $\pm$ 0.18\\
    {\small\textit{s03}} & 0.73 $\pm$ 0.20 & 0.72 $\pm$ 0.10 & 0.68 $\pm$ 0.14 & 0.57 $\pm$ 0.18\\
    {\small\textit{s04}} & 0.70 $\pm$ 0.20 & 0.65 $\pm$ 0.21 & 0.65 $\pm$ 0.26 & 0.54 $\pm$ 0.19\\
    {\small\textit{s05}} & 0.72 $\pm$ 0.19 & 0.67 $\pm$ 0.30 & 0.73 $\pm$ 0.14 & 0.62 $\pm$ 0.18\\
    \bottomrule
  \end{tabular*}
%   \caption{The validation accuracy of arousal and valence models evaluated by 7-fold cross-validation based on two sets of features. \textit{FS} is the set of EEG features selected from fourteen channels as described in Feature Extraction and Selection Section. \textit{T78} is the subset features of extracted EEG features that only related to 
\end{table}

% The last row of Table~\ref{tab:accu} is the match rate of the designated v/a score and the evaluated one in testing experiments, and we call it \textit{Test\_accu}. It validates our strategy of including users' current EEG information for music selection.   for the reason that we could not replay the filtered song under the identical state (EEG information) of the user
Testing accuracy in the last row of Table~\ref{tab:accu} was calculated as the match rate of the designated v/a score and the evaluated v/a score. One of five songs in MAs was selected and presented to the participant. Users' EEG changes every moment, thus songs that were filtered out from MAs were unable to be known whether it was matched or not at that moment. Therefore, the false negative rate and the true negative rate of our system could never be calculated and they both are insignificant to the system function. The true and false positive rates correspond to the \textit{Test\_accu} and ($1-Test\_accu$). The accuracy of recommending new songs is significant for the \textit{Test\_accu} because it's the new observation for the prediction model. And there were ($\#Test\_song - \#New\_song$) old songs that were chosen for each participant in testing experiments, among which some songs had various influences in the training process but matched the user's designation in the testing process. The match rate of old songs is 100\% for participants except for one incorrect prediction for \textit{s02}, which was listened to by \textit{s02} three times and resulted in two different classes of the evaluated v/a scores. It turned out that the system might recommend incorrect old songs with the user's changing tastes and open EEG information. However, it could be solved by updating the prediction model with new data.

\subsection{Emotion Instability}

\begin{table}
\centering
  \caption{Emotion instability of participants}
  \label{tab:randomness}
  \begin{tabular}{c c c c c c}
    \toprule
    % \cmidrule(r){1-6}
    % \hline 
    {\small\textit{}}
    & {\small\textit{s01}}
    & {\small \textit{s02}}
    & {\small \textit{s03}}
    & {\small \textit{s04}}
    & {\small \textit{s05}}\\
    % \hline \hline
    \midrule
    {\small\textit{t\_Arousal}} & 0.24 & 0.15 & 0.22 & {\textbf{0.40}} & {\textbf{0.07}}\\
    {\small \textit{t\_Valence}} & 0.19 & 0.16 & 0.20 & 0.21 & 0.22\\
    % \hline
    % Big-5\_Ex & 19 & 70 & 27 & 41 & 54\\
    {\small \textit{Big-5}} & 0.38 & 0.43 & 0.38 & {\textbf{0.52}} & {\textbf{0.05}}\\
    % \hline
    \bottomrule
  \end{tabular}
%   \caption{The transition values of arousal and valence are represented by \textit{t\_Arousal} and \textit{t\_Valence}, which are calculated by Equation~\ref{equ:2}. Big-5 is the emotion instability score of participants calculated after they performed Five Factor Model (Big-5) test.}
\end{table}
 
The t-score calculated by Equation~\ref{equ:2} for the arousal and valence models are shown as \textit{t\_Arousal} and \textit{t\_Valence} in Table \ref{tab:randomness}. And the collected neuroticism scores from the Big Five Personality Test are shown in \textit{Big-5}. The correlation coefficient between \textit{t\_Arousal} and \textit{Big-5} is 0.808, and between \textit{t\_Valence} and \textit{Big-5} is -0.473. We also found that the participant who had the highest \textit{t\_Arousal} received the highest Big-5 score, and the participant with the lowest \textit{t\_Arousal} received the lowest Big-5 score. Besides, participants $s01$ and $s03$ had the closest \textit{t\_Arousal} and received the same Big-5 score. The p-value between \textit{t\_Arousal} and \textit{Big-5} is 0.073, which is bigger than 0.05 but statistically not random. However, the participants' pool is too small and the songs chosen in experiments are limited, from which we cannot make a general conclusion about the relation between \textit{t\_Valence} and Big-5 score. Here we arguably hypothesize that the t-score can be a referable indicator in emotion-based applications. 

% Hence we are not going to use \textit{t\_Arousal} as a general benchmark of emotion instability at this stage but just an referable indicator. 

% However, with more v/a scores collected with the song, \textit{t\_Arousal} would be more unbiased and representative. the relatively high consistencies between \textit{t\_Arousal} and Big-5 score: thus rejects the null hypothesis that they are unrelated. However, 

\subsection{Results of Real-life Scenarios}
The data collected in training and testing experiments were used to train the emotion variation prediction model of the first day in real-life scenarios. And the model was then updated each day with new data collected, as well as the emotional instability values. There were two experiments for each day; at the beginning of each experiment, participants designated their desired emotion variation direction and stayed still for EEG collection. A list of 10 to 13 songs was selected by steps (2) to (5) of testing experiments iteratively and played. The participants evaluated their v/a scores after listening to each song. 

% in each day is after collecting the data of each day, the prediction model was updated offline for the next day. 

\begin{figure}[tbp!] 
% \centering
    \begin{subfigure}
    \centering
        \includegraphics[width=.32\linewidth,height=.22\linewidth]{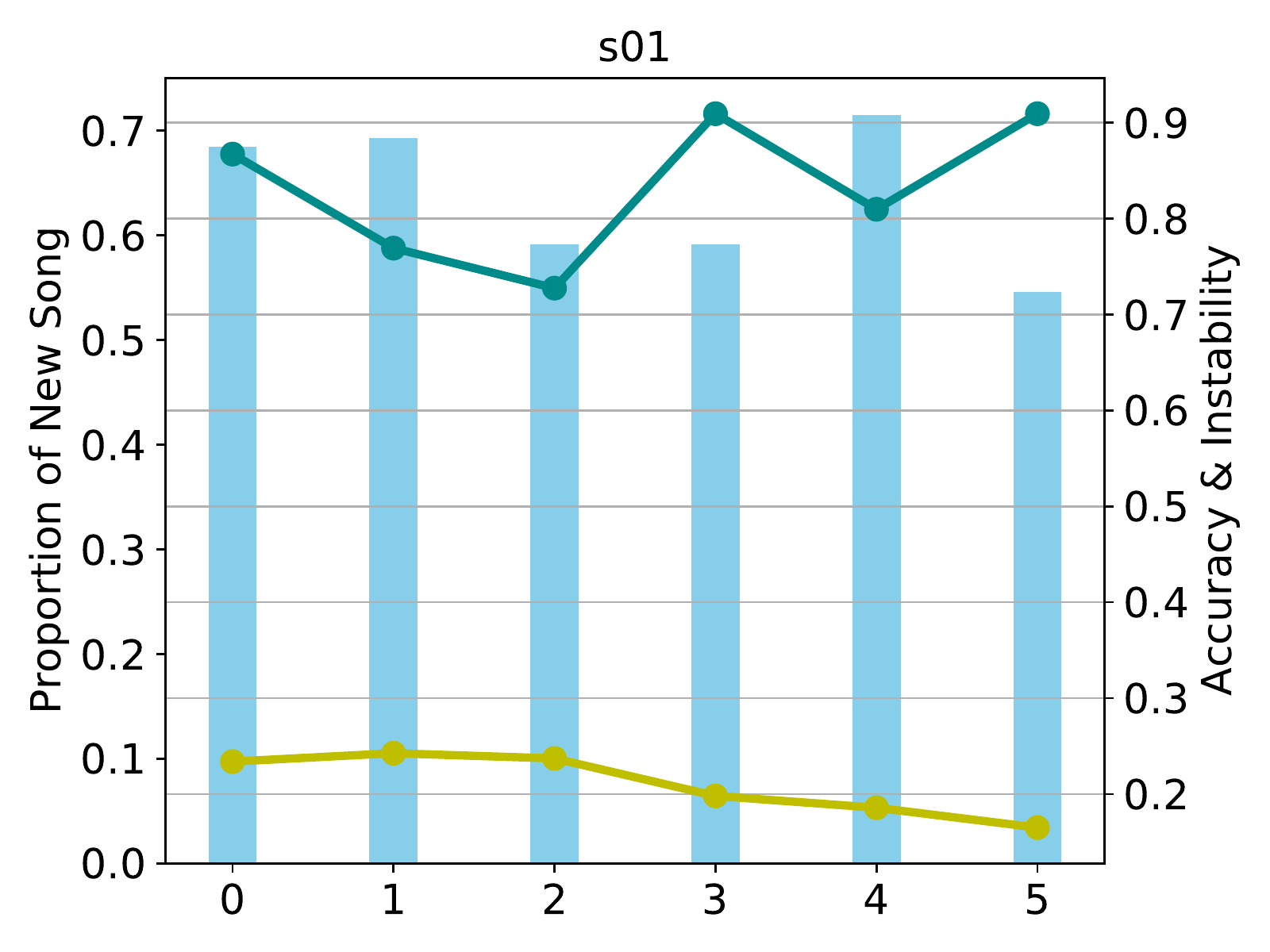}
        % \caption{s01}
    \end{subfigure}%
    ~ 
    \begin{subfigure}
    \centering
        \includegraphics[width=.32\linewidth,height=.22\linewidth]{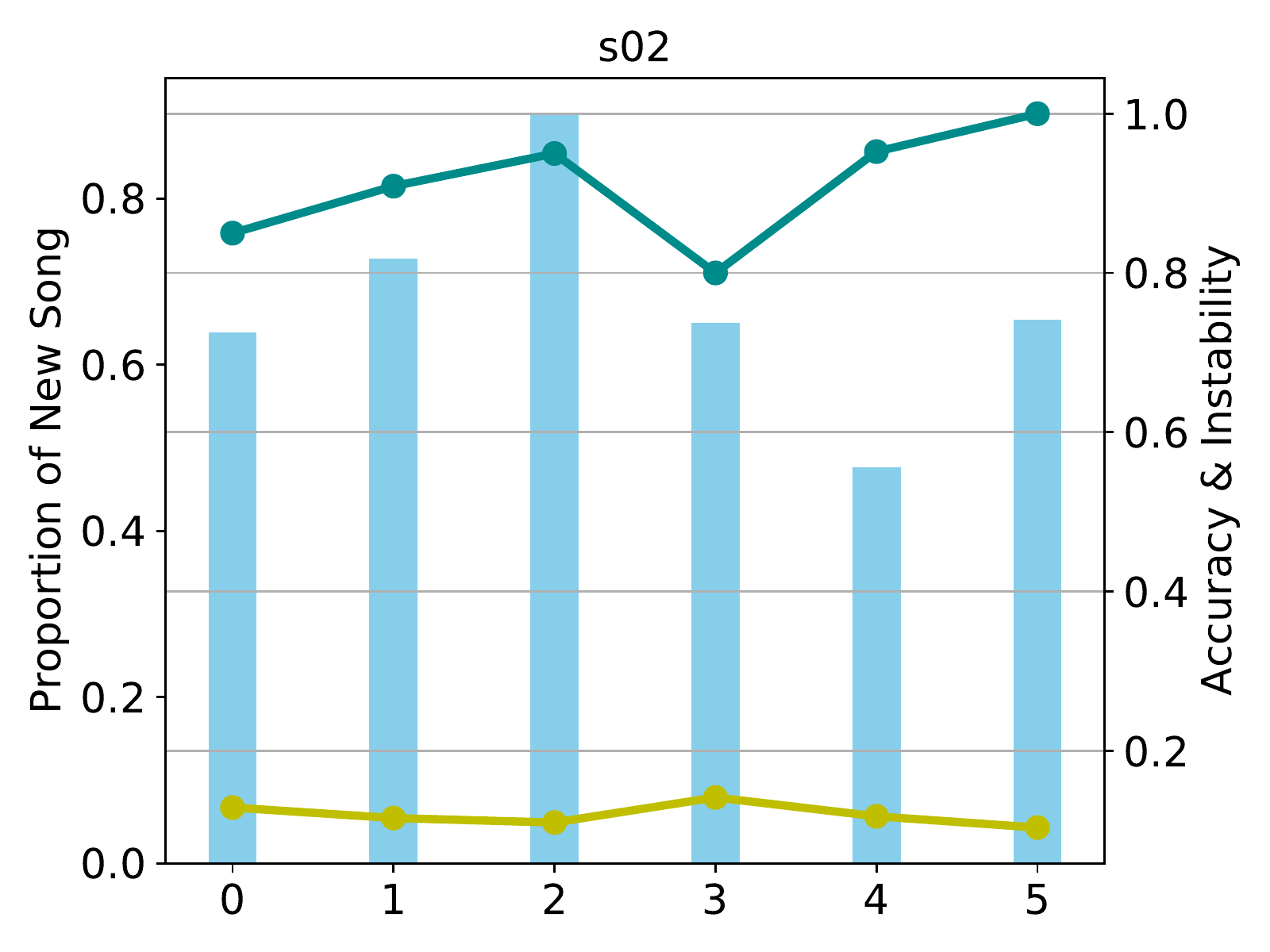}
        % \caption{Lorem ipsum, lorem ipsum,Lorem ipsum, lorem ipsum,Lorem ipsum}
    \end{subfigure}
     ~ 
    \begin{subfigure}
    \centering
        \includegraphics[width=.32\linewidth,height=.22\linewidth]{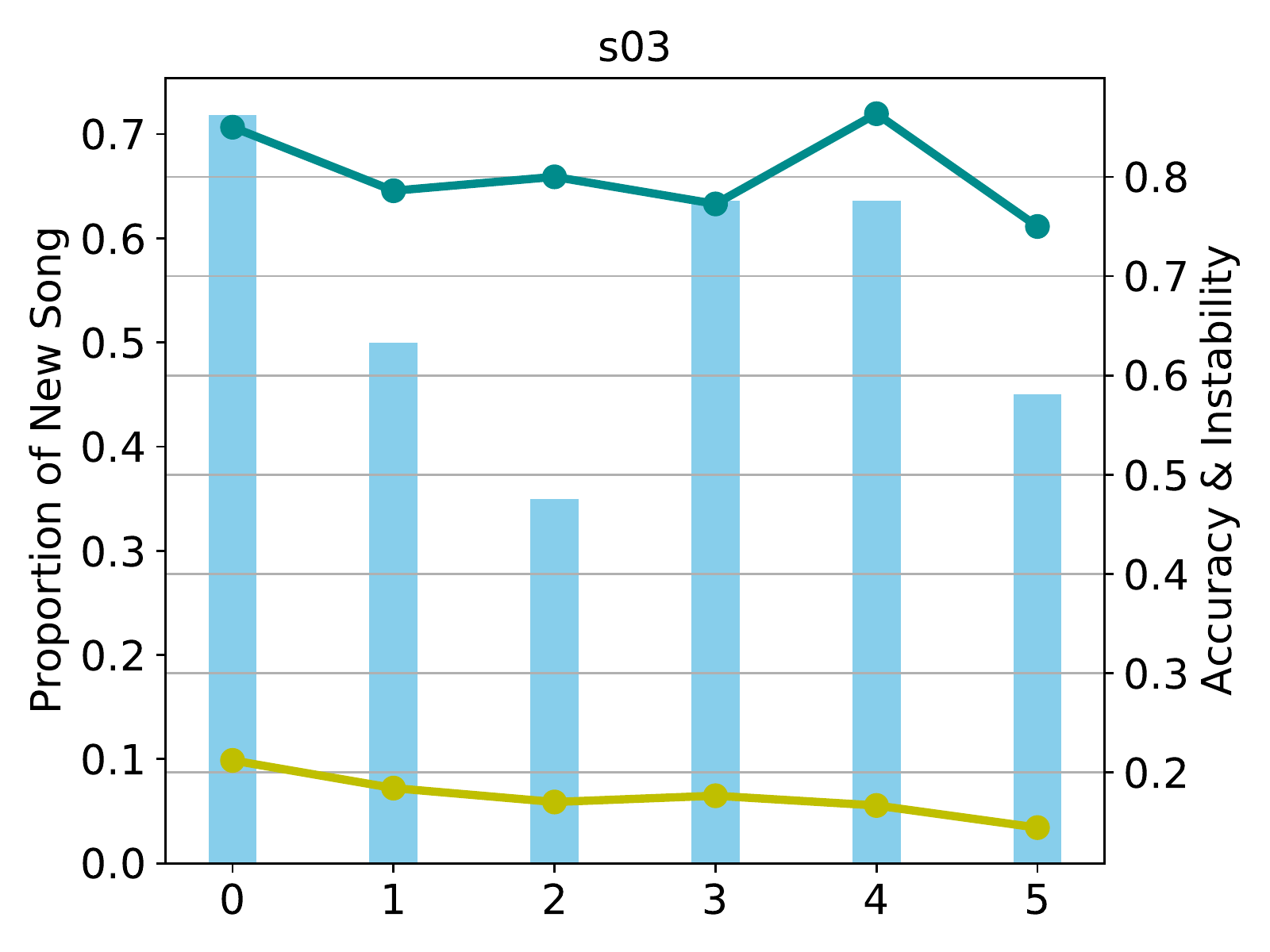}
        % \caption{Lorem ipsum, lorem ipsum,Lorem ipsum, lorem ipsum,Lorem ipsum}
    \end{subfigure}
     ~ 
     \begin{subfigure}
    \centering
        \includegraphics[width=.32\linewidth,height=.22\linewidth]{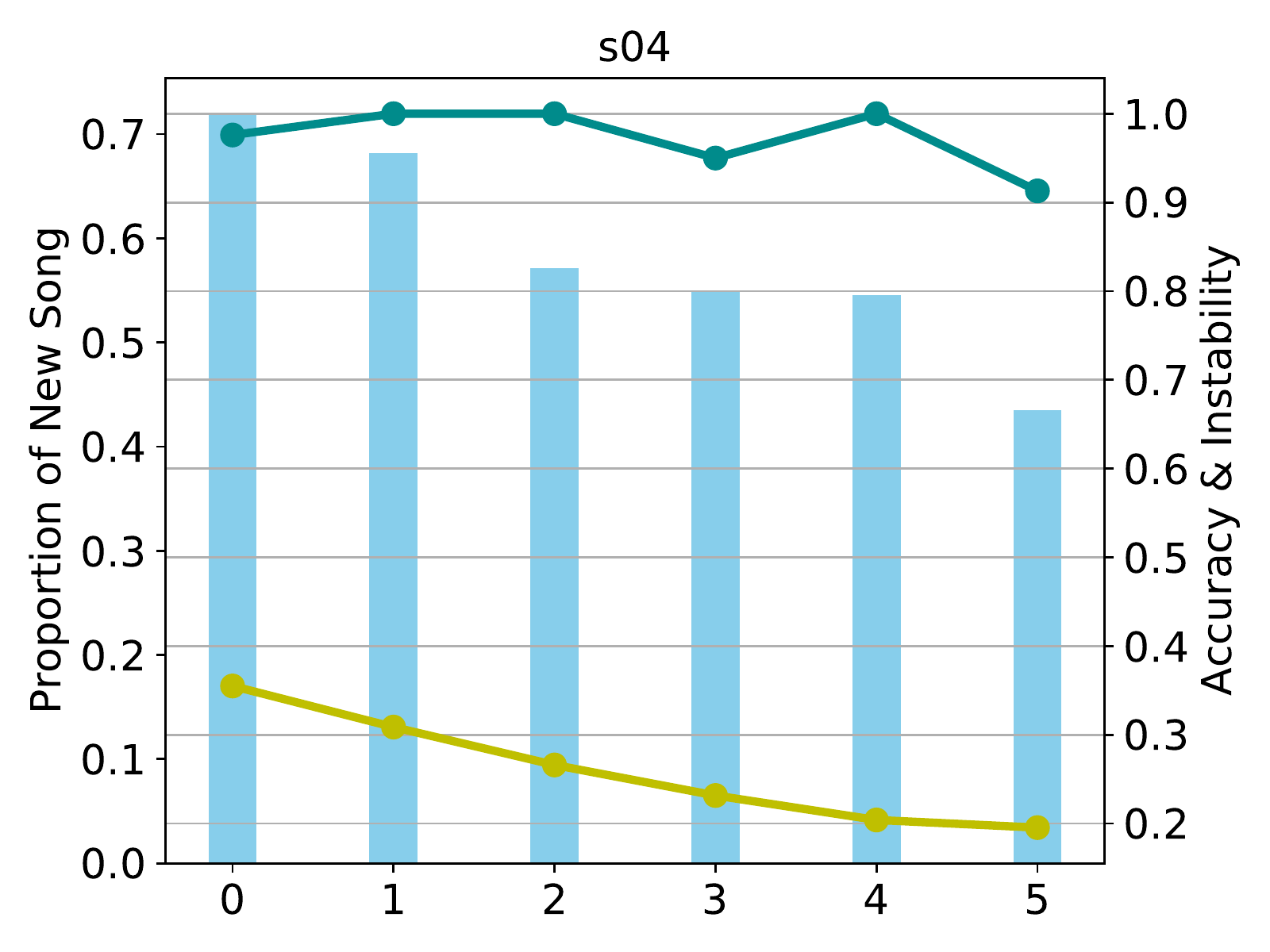}
        % \caption{Lorem ipsum, lorem ipsum,Lorem ipsum, lorem ipsum,Lorem ipsum}
    \end{subfigure}
     ~ 
    \begin{subfigure}
    \centering
        \includegraphics[width=.32\linewidth,height=.22\linewidth]{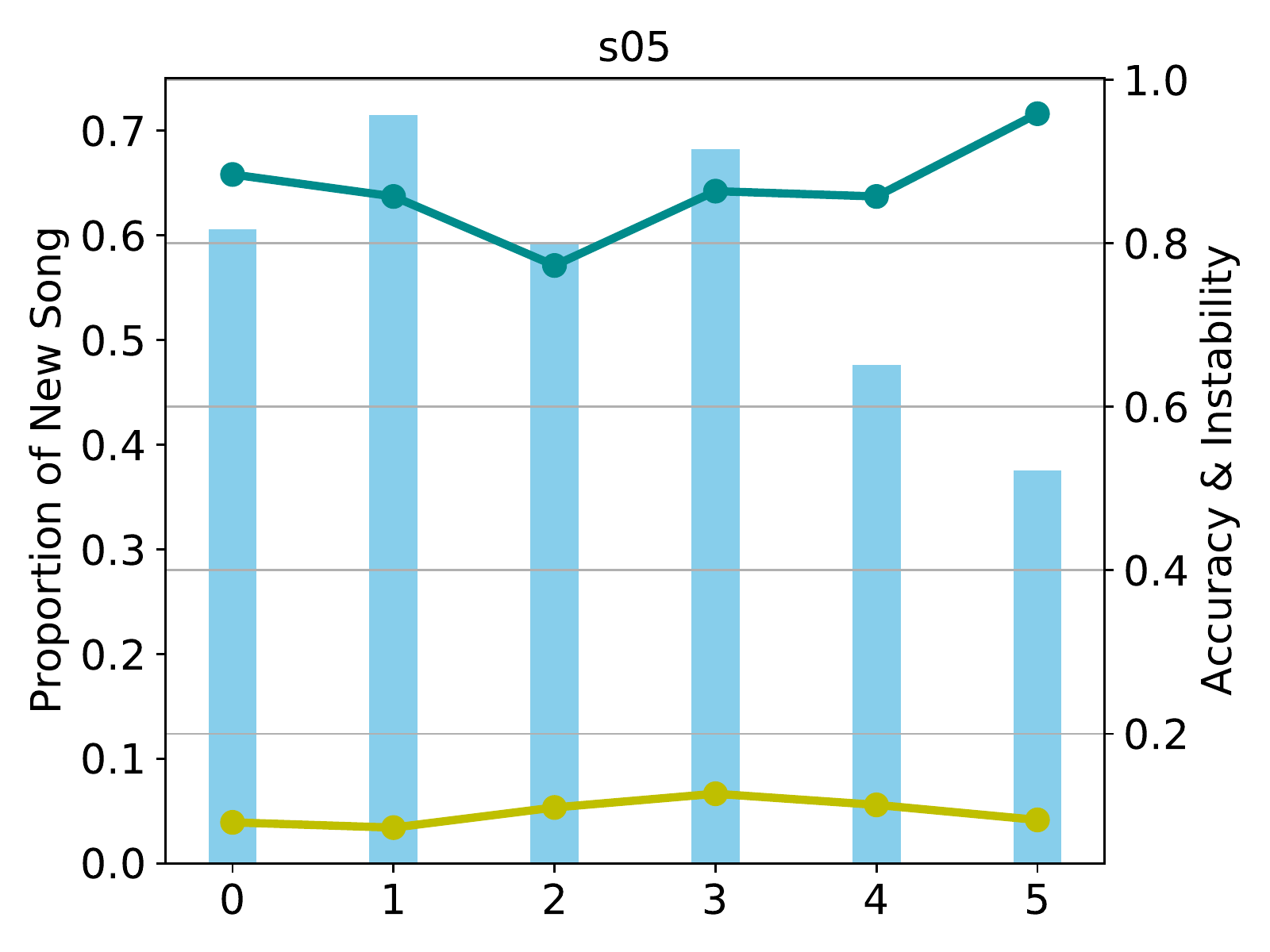}
        % \caption{Lorem ipsum, lorem ipsum,Lorem ipsum, lorem ipsum,Lorem ipsum}
    \end{subfigure}
    ~
    \caption{Testing accuracy (dark cyan line), proportion of new song (blue bar), and emotion instability (yellow line) of each day in real-life scenario. Day 0 is the result of testing experiments.}
    \label{fig:Robu}
    \vspace{-15pt}
\end{figure}

Results were shown in Figure \ref{fig:Robu} for every participant. Day 0 showed the results of testing experiments. The dark cyan line is the testing accuracy of the day. Emotional instability (t-score) was re-calculated with new data and plotted by the yellow line. The proportion of new songs influences the system performance and t-scores besides participants' ever-changing EEG data, thus we also calculated and plotted the ratio of new songs of that day with blue bars for system performance analysis.   
% The proportion of new songs plotted with blue bars, which is the ratio of new songs that never been included for model training among the number of songs listened of the day. and we want to see the stable performance and the convergence of emotion instability in five days.
% which was caused by the system's false prediction of new songs, as well as the confusion of variant scores of old songs.

There are several drops of testing accuracy in Figure \ref{fig:Robu}, including the mismatch of new songs and old songs. When the drop in accuracy was accompanied by the increase in emotional instability (e.g., day 3 of \textit{s02} and \textit{s03}; day 2 of \textit{s05}), it demonstrated that the participant gave different v/a scores to the old song compared with the last time when he listened to it. However, the emotional instability value was mostly averaged by consistent scores of the repeated songs given by participants. The mean t-score of five days for each participant is $[0.206 \pm 0.0009, 0.118 \pm 0.0002, 0.168 \pm 0.0002, 0.241 \pm 0.0017, 0.106 \pm 0.0002]$. All participants have lower instability values as we expected except for a subtle increase for \textit{s05}, who has the lowest emotional instability among them. The small variance of the emotional instability value demonstrates that the songs chosen for participants were not biased, otherwise, it will change dramatically over time. The mean match rate of five days for each participant is $[0.825 \pm 0.0054, 0.922 \pm 0.0046, 0.794 \pm 0.0015, 0.973 \pm 0.0013, 0.862 \pm 0.0035]$. The accuracy is challenged by open EEG data, multi choices of user's designation, and new songs, but remains above $0.8$ and guarantees a satisfactory emotion regulation experience with four choices of emotion variations.
% But the match rate above $0.8$ guarantees a satisfactory emotion regulation experience no matter which direction that users want to change their emotion.
% with  varies insignificantly  The score of emotion instability would be largely biased if it  including new songs or steady increase of accuracy with more training data
% Among four choices that participants can designate their desired emotion variation in real-life scenario,   which means they want to increase their valence

Different from testing experiments where participants were suggested to designate different classes of emotion variation to cover all four classes, they are free to designate their desired emotion variation direction in real-life scenarios and they all designated the first and fourth classes (positive valance changes). It is not surprising that people barely want to change their valence negatively, especially in the experimental environment. But we speculated that people would choose to decrease the valence when they feel hilarious but need to be more ``calm'' and ``serious''. And we didn't build this system by assuming people's specific styles of changing their emotions. Meanwhile, the prediction model is challenged by the unbalanced data collected from real-life scenarios. Even though the regularization parameter is updated with the ratio of binary labels, the prediction accuracy would be degraded with more unbalanced observations. The results from five continuous days of real-life scenarios are insufficient to reveal the long-term performance with more unbalanced data for the missing participants' negative designation of valence. The strategy that makes up for insufficient coverage of all emotional classes and doesn't require users to listen to what they don't designate should be discussed in further studies.

% And it is one advantage of building the system based on emotion variation for users can manage their emotion by direction instead of choosing one specific emotional state, by which the system would be highly biased because people hardly want their emotion to be negative, but just want their emotion changed in that direction. 

% the subtle change of five-day scenario in our situation was reliable but it needed to be proved for a longer period.  ,, for most of the time, people wants to be more positive no matter with higher arousal or lower arousal, especially when they are showing up for experiments in the lab

% \begin{table}
%   \centering
%   \begin{tabular}{c c c c c c}
%     % \toprule
%     \multicolumn{6}{c}{\small{\textbf{Accuracy Comparison of Music Recommenders}}} \\
%     \cmidrule(r){1-6}
%     % \hline 
%     {\small\textit{}}
%     & {\small\textit{Sbj01}}
%     & {\small \textit{Sbj02}}
%     & {\small \textit{Sbj03}}
%     & {\small \textit{Sbj04}}
%     & {\small \textit{Sbj05}}\\
%     % \hline \hline
%     \midrule
%     Music & 0.5 & 0.498 & 0.437 & 0.393 & 0.0708\\
%     EEG\_Music & 0.1930 & 0.1624 & 0.2800 & 0.2390 & 0.465\\
%     % \hline
%     \bottomrule
%   \end{tabular}
%   \caption{Table captions should be placed below the table. We
%     recommend table lines be 1 point, 25\% black. Minimize use of
%     table grid lines.}~\label{tab:mus_fv}
% \end{table}

\subsection{Summary of Questionnaire}
We created a questionnaire on Google Forms with ten questions about users' experience of using music to regulate their emotions. 54 subjects responded, including the five participants in our experiments, in which $38.8\%$ were females, $79.6\%$ of subjects aged from 20 to 29 years, and $18.1\%$ aged above 30 years. For the basic question of how much they like music, from 5 (very much) to 1 (not at all), $50.9\%$ like it very much, $22.6\%$ like it (score 4) and $22.6\%$ chose 3 (moderate). Among all subjects, $86.5\%$ have experienced different feelings towards the same song (without considering familiarity). The question about how well they think the recommendations of a music app could match their tastes \& moods was answered from 5 (very well) to 1 (not at all), $46.7\%$ subjects gave it a 3 and $17.8\%$ were unsatisfied (score 2 \& 1). When asking subjects how annoyed they are when listening to a music piece that they don't like but are recommended by music apps, $86.7\%$ of subjects were annoyed to a varying extent, and the remaining subjects didn't feel annoyed at all. The question about how much they want to try a music recommendation system that can help them regulate their emotional states was answered from 5 (very much) to 1 (not at all). $60.3\%$ of subjects chose 4 and 5, and $13.2\%$ of subjects chose 2 and 1. For the extent of acceptance of two additional EEG sensors bonded with earphones, subjects answered from 5 (very much) to 1 (not at all), $60.4\%$ of subjects chose 4 and 5, and $16.9\%$ chose 1 and 2. 

Most people enjoy music and their feelings are diverse. Thus some of them might feel annoyed when the platform recommends songs based on others' feedback and their old feelings. Our system would particularly benefit users who are willing to regulate emotions through music but don't trust streaming platforms or try to connect with their neural-centric activities. It is not surprising that some people with stable emotions like $s05$ barely have different feelings towards the same song, and they even don't demand emotion regulation. However, besides emotion regulation, the system is open to music and thus can recommend new songs based on users' EEG data. And most people involved in our questionnaire don't refuse to put on a two-electrode EEG device if it could help them with their emotional problems effectively. 

% In addition, all five participants reported satisfactory after experiments --- the music pieces met their demands at that moment.  

% We cannot ignore the annoyance of improper recommendation especially in the process of emotion regulation. 

% =====================================================================
% =====================================================================
\section{Discussions of Limitations and Future Work}

\subsection{Music Dynamics and Influences}
The results of the real-life scenario evaluations revealed the feasibility of recommending a list of songs based on a one-time EEG collection at the beginning. However, the number of songs could be further explored since our choice is based on the number of trials in a training experiment. As people's EEG is dynamically changed by factors like the music they are listening to and the situation they are involved in, the collected EEG information can hardly be representative of more songs. Thus collecting EEG data before selecting each song would be more conservative, which emphasizes the advantage of our system that it is flexible towards people's ever-changing emotional states. Moreover, songs listened to by users in their daily lives are much longer than excerpts we used in experiments. And sometimes, users may enjoy the beginning of the music but start to dislike the remaining part, and vice versa, which diminishes the representation of the averaged features and system performance. In the future, the features that represent the high-level differences between chapters within a song could contribute to a fine-grained solution. Still, users' dynamic self-assessments for a song are necessary, which means a lot more workload for users.

Another concern is related to the impacts of music entrainment, which reflects on users' brainwaves and interferes with their emotional states unconsciously. The impacts of music would last after a song has finished and influence the EEG data collected between two songs. We considered the aftermath of music as trivial in our study, thus we didn't refer to its impacts on this system. But what is reflected on users' brainwaves, and how long will the impacts last are still open questions for studying emotions evoked by music.
% Since users assess their emotion variation based on the whole piece of music, the dynamic emotional state within a song is not considered in our studies. 

% This problem requires more advanced feature engineering and model selection. Moreover, it is undeniable that the context of the song affects the user's emotion. It is hardly to say the context is excluded from the low-level features, but we do not focus on semantic features in this research. 
% And we boldly suppose that the one-time EEG information is possibly valid to recommend songs for hours until users changed their desired emotion variation.  We didn't consider users' dynamic emotion toward a song and averaged the music feature along the time.

% , in which we testify the EEG information within one experiment is similar comparing with other experiments. So that we set the number of songs in a list around the number of songs in a train experiment But in real life, people listen to a complete song, which costs longer time and would involve more or less into the music without lab environment, both conditions would influence users' current state.

\subsection{EEG Interference and Interruptions}
Music features are extracted offline and from a stable source. However, EEG data is collected online and would be easily disturbed and thus mislead the system. The strategy should be considered to make sure that the EEG data collected online is reliable and could be rejected if it is defective. For now, we only employ pre-processing technologies to reject data and channels that are affected by artifacts. With more collected EEG data, it's possible that users' personal EEG profiles could be developed as a reference to evaluate the quality and validity of newly collected data.

Besides, the computational efficiency of a real-time system influences users' experience for possible interruption. And the interruption could be salient and unacceptable when users are focusing on music for emotion regulation. The running time of the music selection in testing experiments is 0.763 seconds (on a 3.4 GHz Intel Core i5 processor) besides the time for EEG collection, in which the EEG processing time by EEGLAB accounts for approximately $80\%$. Decreasing the EEG collection time would significantly relieve the computational workload. Since the slowest component we employed is 8 Hz, a sufficient time for EEG data collection is allowed to be 0.25 seconds theoretically, which should be testified in future work. Another point is that, the acceptance of interruption would be increased when users acknowledge that the system is acquiring EEG signals during that time.

% Features are selected after concatenation based on SBS and they could be different when new data included, thus the quality of EEG data should also be evaluated when choosing final features.   

% Among four choices of desired emotion variation, participants usually designated to the first and forth classes, which biased our collected data. It is reasonable that people want their emotion to be more positive most of time, especially when doing the experiment in the lab, but we speculate that there are situations that people would designate negative variation of valence in real life. For example, when they are at a very excited state but need to be more ``serious'' or ``calm'', then the emotion should be regulated to the negative direction.
% Experiments in more realistic scenarios need to be processed to explore those true experiences of users' emotion regulation.

\subsection{Emotion Models}
There are several clear benefits that users evaluate their emotion variations instead of definitive emotional states, including that it relates the event and users' EEG in one equation, makes it easier for users to report after listening to a song and supports multiple choices of emotion regulation styles. However, since we didn't employ the emotion recognition model, the EEG information didn't reflect any emotional state. Thus we won't return the feedback regarding the user's emotional state based on the collected EEG signals. But we propose that, with sophisticated music information retrieval technologies and users' data involved, the interaction between users and music will provide significant information for neurophysiological studies. Furthermore, four classes of emotion variation choices would be insufficient for users' demands of different emotional change levels. For example, users may want their valence increased substantially but arousal increased slightly. Therefore, emotion variation levels on top of the four basic classes can be further explored for users' personalized, specific emotional demands. Lastly, users' comprehension of their emotions and music tastes may gradually change over time. With more data being collected, the system should take less account of old data and give more weight to the latest data. An online, adaptive preference learning method shall be further developed, instead of the traditional, static classifier. 

% Besides, the recommendation system is an open loop with the designation. A closed loop that detects user's extreme emotional states is more appropriate for emotion regulation, whether by referring to the EEG patterns or by other technical controls. The EEG patterns for extreme emotions could be learnt separately by providing stimulation like intense movie clips. And the technical control can be deve
 
\subsection{System Generalization}
Lastly, the system is highly personalized for participants. However, the experiments for model training before using the online system are time-consuming: more than one hundred trials for each participant in our study. Therefore, the generalization method for new users is strongly demanded in future work. One possible solution is to use advanced classification models like pre-training or fine-tuning. They are built for trained tasks, from which the new customized model can benefit. Another possible solution is to categorize new users into groups by presenting discrepant songs. For users who have similar music tastes or emotion regulation styles, discrepant songs can be used to distinguish their general styles, and then the model can be shared and updated among users in the same category. Furthermore, the system is open to all music pieces, thus it can serve as a second-step music recommender to a user's personal playlist or the playlist recommended by streaming platforms. For example, the playlist could be the one that a user selects in iTunes by mood (like `Sleep'). Then the system will choose the music pieces from the library for the user based on his/her EEG information by setting the designated emotion variation to negative arousal and positive valence, and filtering out the songs that would lead the emotion to higher arousal and cause sleeplessness.

\section{Conclusion}
The interaction between users and music paves the way to comprehending the diverse and ever-changing emotions of people and assists them to regulate emotions in daily life. None of the existing music-based emotion regulation studies has pointed out the limitation of traditional emotion recognition models and come up with the idea that predicts emotion variations, which relate to the song and users' current emotional state. Our system bridges the gap between theoretical studies and real-life applications, and presents the system performance under all four choices of users' desired emotion variations. The robustness of the system is tested with new songs on different days spanning over a period of two months. From users' perspective, the system doesn't return deterministic feedback but follow their wills and emotional states to present stimuli carefully, which assists users to adjust their emotion and leaves space for them to comprehend the emotional changes. We believe that users can get benefits from such decent and user-oriented interaction, which really considers their variant emotions during the process. 

\section*{Acknowledgment}
This material is based upon work supported by the National Science Foundation under Grant No. CNS-1840790. Any opinions, findings, and conclusions or recommendations expressed in this material are those of the author(s) and do not necessarily reflect the views of the National Science Foundation.

% \begin{figure}
% 	\centering
% 	\fbox{\rule[-.5cm]{4cm}{4cm} \rule[-.5cm]{4cm}{0cm}}
% 	\caption{Sample figure caption.}
% 	\label{fig:fig1}
% \end{figure}

% \subsection{Tables}
% See awesome Table~\ref{tab:table}.

% The documentation for \verb+booktabs+ (`Publication quality tables in LaTeX') is available from:
% \begin{center}
% 	\url{https://www.ctan.org/pkg/booktabs}
% \end{center}

\bibliographystyle{unsrtnat}
\bibliography{references}  %%% Uncomment this line and comment out the ``thebibliography'' section below to use the external .bib file (using bibtex) .





% \end{thebibliography}

\end{document}